\documentclass[twocolumn,aps,amsmath,amssymb,floatfix,prb]{revtex4}
\usepackage{graphicx}
\usepackage{dcolumn}
\usepackage{bm}
\usepackage{amsfonts}
\usepackage{graphicx}
\usepackage{dcolumn}
\usepackage{bm}
\usepackage{amsmath}
\usepackage{amssymb}

\newcommand{\vk}{{\bf k}}
\newcommand{\ve}{{\varepsilon}}
\newcommand{\vek}{{\varepsilon_k}}

\newcommand{\vq}{{\bf q}}

\begin{document}

\title{Linear-in-$T$ resistivity in dilute metals: A Fermi liquid perspective}
\author{E. H. Hwang$^1$ and S. Das Sarma$^2$}
\affiliation{
$^1$SKKU Advanced Institute of Nanotechnology and Department of
Nano Engineering, Sungkyunkwan  University, Suwon, 16419, Korea \\
$^2$Condensed Matter Theory Center and Joint Quantum Institute, 
Department of Physics, University of Maryland, College Park,
Maryland  20742-4111 U.S.A. \\
}



\begin{abstract}
We consider a short-range  deformation potential scattering model of electron-acoustic phonon interaction to calculate the resistivity of an ideal metal (i.e. no other scattering mechanism except for acoustic phonon scattering) as a function of temperature ($T$) and electron density ($n$).  The resistivity calculation is based on the Boltzmann transport theory within the relaxation time approximation in the nearly free electron single band approximation.  We consider both 3D metals and 2D metals, and focus on the dilute limit, i.e., low effective metallic carrier density (and hence low effective Fermi wavenumber $k_F$) of the system.  The main findings are : (1) a phonon scattering induced linear-in-$T$ resistivity could persist to arbitrarily low temperatures in the dilute limit independent of the Debye temperature ($T_D$) although eventually the low-$T$ resistivity turns over to the expected Bloch-Gr\"{u}neisen (BG) behavior with $T^5$ ($T^4$)  dependence, in 3D (2D) respectively, with the crossover temperature, $T_{BG}$, from the linear-in-$T$  to the  BG behavior, being proportional to the Fermi momentum, is small in the dilute limit; (2) because of low values of $n$, the phonon-induced resistivity could be very high in the system, orders of magnitude above the corresponding room temperature resistivity of ordinary metals; (3) the resistivity shows an intrinsic saturation effect at very high temperatures (for $T>T_D$), and in fact, weakly decreases with increasing $T$ above a high crossover temperature with this crossover being dependent on both $T_D$ and $n$ in a non-universal manner --  this high-temperature crossover is not directly connected with the Mott-Ioffe-Regel limit and is a reflection of phonon phase space restriction.  We discuss the qualitative trends in the resistivity as a function of temperature, density, phonon velocity, and system dimensionality. We also provide `high-temperature' linear-in-$T$ resistivity results for 2D and 3D Dirac materials. Our work brings out the universal features of phonon-induced transport in dilute metals, and we comment on possible implications of our results for strange metals, emphasizing that the mere observation of a linear-in-$T$ metallic resistivity at low temperatures or a very high metallic resistivity at high temperatures is not necessarily a reason to invoke an underlying quantum critical strange metal behavior.  Dilute metals may very well have highly unusual (compared with normal metals) transport properties arising from quantitative, but not qualitatively new, underlying physics.
We discuss the temperature variation of the effective transport scattering rate showing that for reasonable parameters the scattering rate could be below or above $k_BT$, and in particular, purely coincidentally, the calculated scattering rate happens to be $k_BT$ in normal metals with no implications whatsoever for the so-called Planckian behavior.
Our work manifestly establishes that an apparent Planckian dissipative behavior could arise from the usual electron-phonon interaction without implying any strange metallicity or a failure of the quasiparticle paradigm in contrast to recent claims.
\end{abstract}

\maketitle


\section{Introduction}

Electrical transport is a key property of electronic materials, distinguishing between metals and insulators: Metals conduct electricity at $T=0$ (have nonzero conductivity) whereas insulators do not (zero conductivity  $\sigma=0$ at $T=0$).  Even at room temperatures the resistivity $\rho$ ($=1/\sigma$) ratio between a good metal (e.g., Ag, Cu) and a good insulator (e.g., Teflon, wood) could be as large as $10^{20}$ to $10^{30}$ making it one of the largest dimensionless numbers occurring  naturally in materials science.  Often metals (insulators) are defined loosely as materials whose resistivity increases (decreases) with increasing temperature. 
Many of the major discoveries and advances in condensed matter physics arose from the measurement of electrical transport properties (e.g. superconductivity, quantum Hall effects, giant magnetoresistance, localization, Kondo effect).  
Understanding dc-transport properties of materials is among the oldest problems in solid state physics (e.g., Wiedemann-Franz law, Drude theory)\cite{Ziman,Ziman2,ashcroft}. Investigations of various types of metal-insulator transitions (e.g., Anderson localization, Mott transition)\cite{localization0,localization,localization2} by tuning system parameters (e.g., density, pressure, temperature, magnetic field, doping, disorder, chemical potential) have been (and still are) among the most-researched topics in condensed matter physics.  Studying dc-transport in great details remains the essential aspect in the elucidation of the properties of any new material, and the understanding of metallic conductivity as a direct manifestation of the existence of a Fermi surface is one of the great fundamental advances in quantum physics.

In this context, the concept of a `strange metal' (often the terminology `bad metal' is used to convey the same idea -- we use the terminology `strange metal' and `strange metallicity' in the current work) has emerged as an important new idea over the last 30 years.  Although no strict formal definition exists, strange metallicity involves a set of unusual (as compared with simple normal metals such as Al and Cu) transport properties such as  a linear-in-$T$ resistivity over a large temperature range and a very high metallic resistivity at room temperatures much larger than the typical metallic value of $1-10 \mu \Omega \cdot$cm occasionally going above the so-called Mott-Ioffe-Regel (MIR) limit\cite{ioffe,mott}  of 150 $\mu \Omega\cdot$cm.  It is often thought that strange metals, in contrast to `good or normal metals' (e.g. usual elemental metals such as K, Li, Cu, Al), do not obey the Landau Fermi liquid quasiparticle paradigm and are consequently examples of correlation-driven non-Fermi liquids.  
The concept of strange metals originally arose in the context of normal state transport properties of high-temperature cuprate superconductors,\cite{kiverson,bad} which often reflect a linear-in-$T$ resistivity approaching $\sim$m$\Omega \cdot$cm at room temperatures, but has now been extended to large classes of metallic compounds (e.g., cuprates\cite{cuprate}, ruthenates\cite{ruthenate}, titanates\cite{titanate}, manganites\cite{manganite}, heavy fermions\cite{heavy}), where linear-in-$T$ transport along with very high metallic resistivity manifest (at least in some regimes of the experimental parameter space).  In fact, converting the transport relaxation time $\tau$ into an equivalent temperature $\hbar/k_B \tau$ it has even been argued that strange metals are often in the so-called Planckian limit \cite{Bruin,zaanen}
where the transport relaxation time is given simply by the measurement temperature $T= \hbar/k_B \tau$ -- presumably this description applies only at finite temperatures (and an implicit subtraction of any extrinsic impurity contribution to the resistivity is perhaps implied although often not explicitly stated).  Strange metallicity is often thought to be connected with a hidden quantum critical point, and the thermal excitations in the quantum critical `fan' region above the critical point are thought to be responsible for the strange metallic behavior although explicit calculations starting from a realistic microscopic Hamiltonian incorporating both quantum criticality and the resultant strange metallicity are rare.  Basically, `strange metals' are metals with resistivity high enough that they should almost qualify as insulators (hence `strange') in addition to having a linear-in-$T$ resistivity behavior over a large range of temperature.
In fact, the precise definition of strange metallicity is a little vague, with different experimental groups attributing different characteristics to the definition, but very high and linear-in-$T$ resistivity seem to be a common theme.

Strange metallicity is the indirect motivation for the current work although we will not make contact with experimental data in any strongly correlated materials directly.  We will, however, show that some (but, by no means all) aspects of the putative strange metallic behavior may arise, both in three-dimensional (3D) and two-dimensional (2D) systems, simply by virtue of strong electron-phonon scattering and low carrier density, i.e., in dilute metals.  
The carrier effective mass, if it is heavy, also plays a role in our theory for strange metallicity -- basically, we find that large values of $m/n$, where $m$ ($n$) are carrier effective mass (density), generically lead to strange metal transport properties through electron-phonon interaction.
In particular, a linear-in-$T$ resistivity arising purely from electron-phonon scattering could persist to arbitrarily low temperatures in dilute 2D and 3D metals provided that the system is sufficiently dilute (i.e., very small $n$).  In addition, the room temperature intrinsic (i.e., phonon-induced) resistivity of such dilute 2D and 3D metals could be very large by virtue of the low carrier density even for reasonable and realistic electron-phonon coupling strengths.  We will thus establish that low carrier density coupled with standard electron-phonon interaction is capable of producing resistive behavior mimicking strange metallicity without incorporating any strong correlation or electron interaction or quantum criticality effects.  
Thus, `strange' metallicity could arise from ordinary electron-phonon interaction without invoking any exotic mechanism.  In particular, we show that the phonon-scattering induced linear-in-$T$ metallic resistivity could easily surpass the Planckian limit, thus seriously calling into question any deep significance associated with the transport scattering rate becoming equal to (or even larger than) the temperature of the system.

With the above background, we study theoretically in this paper in great depth a classic problem in metallic dc-transport, namely, noninteracting electrons confined to a single parabolic band, characterized by an effective mass $m$ and a carrier density $n$, interacting with acoustic phonons of the underlying lattice in the continuum approximation, giving rise to a phonon scattering induced metallic electrical resistivity $\rho(n,T)$.  
(We also provide some results for Dirac-like linear band dispersion as appropriate for graphene
although this is not the focus of this work.)
The calculation is carried out entirely using the semiclassical Boltzmann transport theory \cite{Ziman2} with the scattering rates calculated using the quantum Fermi's golden rule approximation and the appropriate thermal averages done using the Fermi distribution function for the carriers.  
The electron-phonon coupling constant (i.e., the deformation potential coupling strength) is kept small enough so that the leading order scattering theory using the golden rule is adequate with higher order terms being negligibly small.
We ignore disorder and electron-electron interaction effects (although some effects of electron-electron interaction could in principle be subsumed in the effective mass $m$ and the electron-phonon interaction strength $D$ through the standard many-body  Fermi liquid renormalization). Because of the neglect of disorder effects (i.e., no extrinsic resistivity) the system we study is a perfect metal at $T=0$ with  $\rho(T=0)=0$.  
The electron-phonon interaction strength $D$ (often called the deformation potential coupling constant, which depends on the material and is typically $D \sim1-50$ eV) decides the overall magnitude of the resistivity (to be more precise $\rho \sim D^2$), but the $(n,T)$ functional dependence of $\rho$ is determined by the interplay of three characteristic temperature scales: the Fermi temperature $(T_F)$, the Bloch-Gr\"{u}neisen temperature ($T_{BG}$), and the Debye temperature ($T_D$) given by: $k_BT_F=E_F/k_B$, $k_BT_{BG}=2\hbar v_s k_F$, $k_BT_D=(\hbar v_s)  (6\pi^2 N)^{1/3}$, where $v_s$ is the sound (i.e., acoustic phonon) velocity, $k_F$ is the electronic Fermi wave number (and $E_F=(\hbar k_F)^2/2m$ is the Fermi energy), and $N$ is the atomic number density.  We take $T_D$ as a parameter, but $T_F$ (and $E_F$) and $T_{BG}$ are calculated using the carrier density $n$, electron effective mass $m$, and phonon velocity $v_s$.  Note that the physics is determined by two velocities $v_s$ (which is a constant parameter independent of carrier density) and $v_F =\hbar k_F/m$ with $k_F \sim n^{1/d}$ where $d$ ($=2$ or 3) is the system dimensionality.  Overall, there are seven independent parameters in the problem ($d$, $D$, $v_s$, $T_D$, $m$, $n$, $T$), two of which depend on the carriers ($m$, $n$) and two on the lattice ($v_s$, $T_D$) whereas $D$ is the interaction strength coupling the electron-phonon system.  The other two parameters $d$ (dimensionality) and $T$ (temperature) describe the combined system as a whole -- we assume that the electrons and phonons have the same dimensionality and are in equilibrium with each other being at the same temperature $T$.  All of these approximations are standard in the context of electron-phonon transport theories in ordinary metals.
In particular, the use of Boltzmann transport theory coupled with the golden rule relaxation time approximation is standard in the calculation of metallic resistivity arising from electron-phonon scattering. \cite{Ziman,Ziman2,ashcroft}

The system we study is therefore as normal or ordinary (i.e., non-strange) as it can be, and is a textbook model of a normal Fermi liquid. The only difference is that, in contrast to normal metals which have fixed (and rather high) density, $n \sim 10^{23}$ per cm$^3$, we assume the metallic density to be a variable and we vary the temperature over a wide range.  We also carry out the transport calculation without imposing $T_D$ to be the absolute energy cut off for the phonons (as is done in textbooks)-- we take into account the fact that at low carrier densities the phonon cut off is really $T_{BG}$ as long as $T_{BG}<T_D$.
Our definition of a dilute metal is that the carrier density $n$ is low enough for $T_D > T_{BG}  \sim n^{1/d}$. This implies a low density depending on the phonon Debye temperature, which is purely a lattice property of the system.  Normal metals, by contrast, have $T_D<T_{BG}$ since normal metals have very high electron density ($ \sim 10^{23}$ cm$^{-3}$).  Doped semiconductors (both 2D and 3D) are dilute metals by our definition since the typical equivalent 3D carrier density in semiconductors is $<10^{20}$ cm$^{-3}$.  Most highly correlated materials manifesting strange metal behavior are also dilute by our definition as they tend to have $T_{BG}<T_D$ with $n \sim 10^{21}$ cm$^{-3}$. The important point here is that the phonon induced linear-in-$T$ resistivity persists down to $T \sim T_{BG}/$3 in dilute metals, which could be very low depending on the actual carrier density, since $T_{BG} \sim k_F \sim n^{1/d}$.

Given such a large number of parameters, including two continuous parameters $n$ and $T$, the possibilities for $\rho(n,T)$ are huge depending on the values of the other 5 parameters.  We will in fact present density and temperature dependent resistivity results for various values of $m$ so that the quantitative effect of $T_F \sim 1/m$ can be discerned, but we will stick to fixed values of the lattice parameters $D$, $v_s$, and $T_D$ since no new qualitative physics arises by varying these parameters.  We will compare results for $d=2$ and 3 since both 2D and 3D materials are of interest in the context of strange metal physics as many of the strange metals (e.g. cuprates) are essentially 2D in nature because of their highly anisotropic effective mass. The most important question to be addressed in the work is the ($T$, $n$) regime over which the resistivity is linear in $T$ as well as the absolute values of the calculated resistivity.  Both of these issues of course depend on the choice of the system parameters, and we choose our lattice parameters using GaAs as the guide.  The rationale for using GaAs is that experimental transport data already exist in metallic  2D GaAs structures with $\rho (T)$ being linear down to 100 mK (or lower) both for electrons and holes.  Thus, 2D GaAs is one system where the physics being elaborated in the current work has already manifested itself, making it reasonable that we use GaAs parameters as the starting point -- an obvious advantage being that the theoretical results agree with the experimental data in the extensively studied high-mobility 2D GaAs semiconductor structures.  We do, however, change $m$ and $n$ enough to cover the parameter regime of interest for cuprates and related systems.  Fortunately, the phonon parameters do not typically vary much (by less than a factor of 2) among different electronic materials, and the deformation potential coupling strength $D$ is simply an overall scale of the resistivity, thus varying $D$ only changes the resistivity by $D^2$.
Thus, the precise value of $D$ ($=12$ eV) being used in the current work is not particularly germane since results for other materials can simply be obtained by multiplying our calculated resistivity values by ($D_{\rm new}/12)^2$ where $D_{\rm new}$ is the deformation potential coupling in electron-Volts in the new material (of course the other parameters such as {\it m, n, T}, and $v_s$ must also be the same).

The rest of the paper is organized as follows.  We provide the basic transport theory for calculating the resistivity arising from electron-phonon coupling in Sec.~II, where we also discuss the details of the various parameters used in our calculations.  We give detailed analytical and numerical results for  $\rho(T, n)$ for 2D metals also in Sec.~II and for 3D metals in Sec.~III, keeping all the parameters consistent with each other.  
In sec. IV, we briefly discuss the linear-in-$T$ resistivity in Dirac materials (e.g. graphene in 2D) arising from electron-phonon interaction.
We provide a discussion of our results in Sec.~V, where we also compare 2D and 3D resistivity results, and comment on strange metallicity in the context of our results.
We conclude in Sec.~VI. 

\section{Theory and 2D resistivity}

In this section we describe the Boltzmann theory for electronic resistivity ($\rho$) due to the longitudinal acoustic phonon scattering. We consider the deformation potential scattering which arises from quasistatic deformation of the lattice. 
Within Boltzmann transport theory by averaging over energy at finite temperatures, we obtain the conductivity ($\sigma = 1/\rho$) \cite{Ziman2,ashcroft}
\begin{equation}
\sigma = \frac{e^2}{d}\int d\ve N(\ve) {v}_{\vk}^2 \tau(\ve) \left (- \frac{\partial f(\ve)}{\partial \ve} \right ),
\label{eq:sigma}
\end{equation}
where $d$ represents the dimension of the system,  $N(\ve)$ is the density of states,
$\ve = \hbar^2 k^2/2m$ the noninteracting kinetic energy, $\vk$ the wave vector, $v_{\vk} = \hbar \vk/m$  the carrier velocity, $\tau(\ve)$ the transport relaxation time, and $f(\ve)$ is the Fermi distribution function. At $T=0$, $f(\ve)$ is a step function at the Fermi energy ($E_F$), and we have the usual conductivity formula
(for parabolic metals with a carrier density $n$)
\begin{equation}
\sigma = \frac{e^2 v_F^2}{d} N(E_F) \tau(E_F) = \frac{n e^2 \tau(E_F)}{m}.
\end{equation}

Taking $\vk$ and $\vk'$ to denote the electron wave vectors before and after scattering by a phonon, respectively, the energy dependent relaxation time [$\tau(\ve_{\vk})$] is given by\cite{Ziman2}
\begin{equation}
\frac{1}{\tau(\ve_{\vk})} = \sum_{\vk'}(1-\cos\theta_{\vk \vk'}) W_{\vk
  \vk'}\frac{1 - f(\ve_{\vk})}{1-f(\ve_{\vk'})}
\label{tau_3d}
\end{equation}
where $\theta_{\vk \vk'}$ is the scattering angle between $\vk$ and $\vk'$, and
$W_{\vk \vk'}$ is the transition rate from the initial state with 
momentum $\vk$ to the final $\vk'$ state. 
When we consider the relaxation time due to deformation potential (DP) coupled acoustic phonon mode, 
then the transition rate has the form \cite{price,KawamuraPRB1990}
\begin{equation}
W_{\vk \vk'}=\frac{2\pi}{\hbar}|C(\vq)|^2
\Delta(\varepsilon,\varepsilon')   
\label{wkkp}
\end{equation}
where $\vq = \vk-\vk'$ and $|C(\vq)|^2$ is the matrix element for scattering by acoustic phonon. The matrix element $|C(\vq)|^2$ for the deformation potential coupling is given by
\begin{equation}
|C(\vq)|^2 = \frac{D^2\hbar q}{2\rho_0 v_{s}},
\end{equation}
where $D$ is the deformation potential and $\rho_0$ is the mass density. 
In Eq.~(\ref{wkkp}) the factor $\Delta(\ve,\ve')$ is given by
\begin{equation}
\Delta(\ve,\ve') = N_q \delta(\ve-\ve'+\hbar\omega_{\vq}) + (N_q + 1)
\delta(\ve-\ve'-\hbar\omega_{\vq}),
\label{delta}
\end{equation}
where  $\omega_{\vq}=v_{s} \vq$ is the acoustic phonon energy with $v_{s}$ being the phonon or sound velocity,  $\ve = \vek$, $\ve' = \ve_{k'}$,  and
$N_q$ is the bosonic phonon occupation number
\begin{equation}
N_q = \frac{1}{\exp(\beta \omega_{\vq}) -1},
\label{n_q}
\end{equation}
where $\beta = k_B T$.
The first (second) term in Eq.~(\ref{delta}) corresponds to the
absorption (emission) of an acoustic phonon of wave vector $\vq = \vk-\vk'$.
Note that the matrix element $|C(\vq)|^2$ is independent of the phonon
occupation numbers.

\begin{figure*}[t]
\vspace{10pt}%
\includegraphics[width=.8\linewidth]{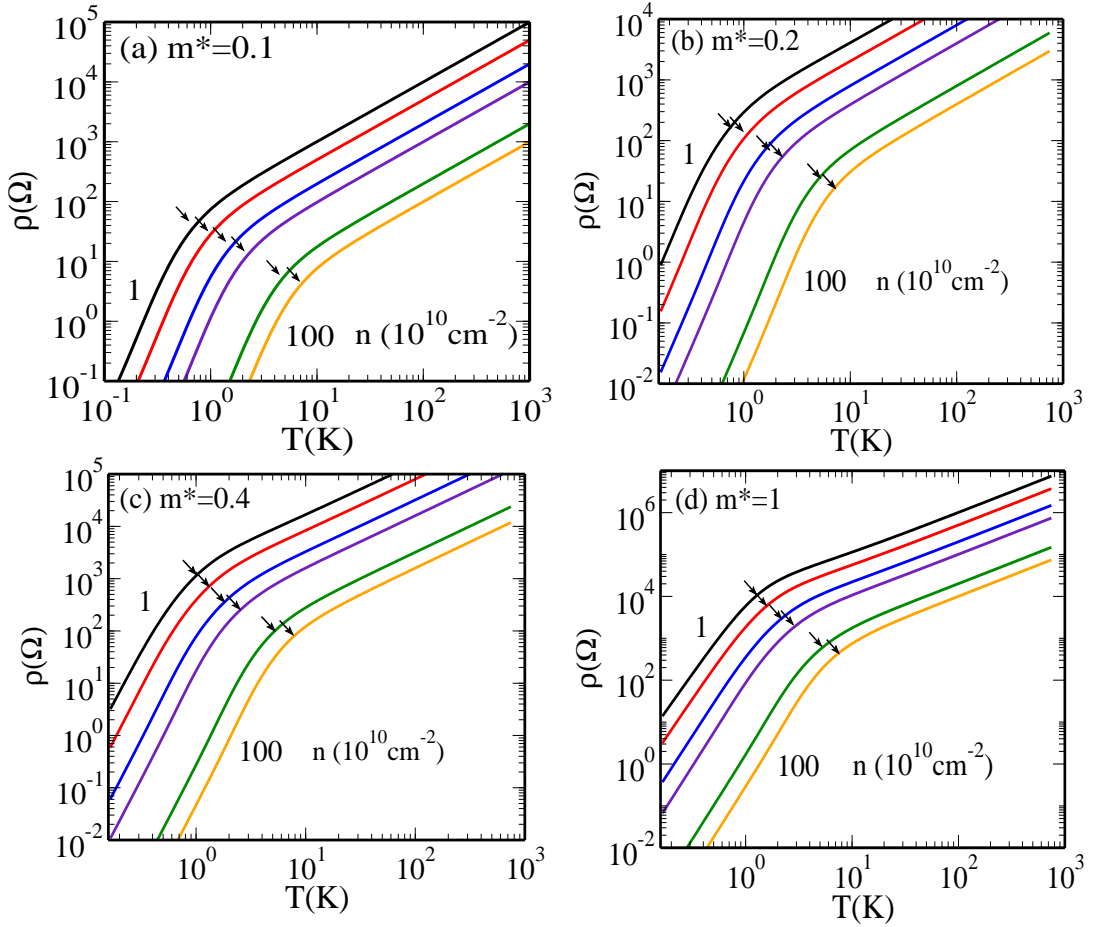}
\caption{
Calculated 2D resistivity limited by the acoustic phonon scattering as a function of temperature for various effective masses and electron densities  ($n\times 10^{10}$cm$^{-2}$, where $n=1, \; 2, \; 5, \; 10, \; 50, \; 100$ from top to bottom).
Here $m^*=m/m_e$ with electron bare mass $m_e$ and the small arrows indicate the crossover temperatures, $T_c$, in which the resistivity ($\rho \propto T^{\alpha}$) changes the power law from high ($T^4$) to linear ($T$) behavior as temperature increases.
The parameters corresponding to GaAs ($v_s = 5.14 \times 10^5$ cm/s, $D=12$ eV, $\rho_{3D} = 5.32$ g/cm$^3$) and the characteristic width of a 2D system $w = 10$ nm are used. With these parameters the BG and Fermi temperatures are given by $T_{BG}=1.97 \sqrt{\tilde{n}}$ and $T_F = 0.28 \tilde{n}/m^*$, where $\tilde{n}$ is the density measured in unit of $10^{10}$ cm$^{-2}$.
}
\label{fig:rho_2d_m1}
\end{figure*}

We note that we are using Fermi's golden rule or the Born approximation to calculate the relaxation rate, which is standard for transport calculations.  For our problem, this leading order approximation for electron-phonon scattering is justified by the weak coupling nature of electron-phonon interaction in our systems of interest.  In particular, the dimensionless electron-phonon coupling parameter (or equivalently, the Eliashberg coupling constant), $\lambda$ ($\sim D^2$),\cite{grimvall}
is rather small ($\ll 1$), validating the leading order scattering theory.  Most, if not all, transport theory calculations for electron-phonon scattering are carried out within this leading order approximation, with the resistivity being proportional to $D^2 \sim \lambda$.


In 2D the electron density $n$ is given by $n = \int N(\ve)f(\ve) d\ve$ with $N(\ve)=m/\pi\hbar^2$. Then Eq.~(\ref{eq:sigma}) can be written as $\sigma = ne^2 \langle \tau \rangle /m$, where $\langle \tau \rangle$ is the energy averaged scattering time \cite{ando,hwangRMP}
\begin{equation}
\langle \tau \rangle =  \frac{ \int d\ve \ve \tau(\ve) \left (-\frac{\partial f(\ve)}{\partial \ve} \right ) }  {
\int d\ve \ve \left (-\frac{\partial f(\ve)}{\partial \ve} \right ) }.
\label{eq:tau_2d}
\end{equation}
Using 2D wave vectors (\vk, $\vk'$, \vq) in Eqs.~(\ref{tau_3d})--(\ref{n_q}) and $\rho_0=\rho_{3D} w$ for 2D case, where $\rho_{3D}$ is the 3D atomic mass density of a material and $w$ is the characteristic width of the 2D system
('$w$' should be considered the typical thickness of the 2D layer or an appropriate interlayer separation for layered materials such as cuprates),
we express the scattering time in 2D as
\begin{equation}
\frac{1}{\tau(\vek)} =\frac{1}{\tau_a} + \frac{1}{\tau_e},
\end{equation}
where $\tau_a$ and $\tau_e$ are the scattering time corresponding to the absorption and emission of phonons, respectively, and they are given by 
\begin{equation}
\frac{1}{\tau_a} = A \int_{0}^{\theta_{m}}d\theta (1-\cos\theta) {q} N_q \frac{1-f(\ve + \hbar \omega_q)}{1-f(\ve)},
\end{equation}
\begin{equation}
\frac{1}{\tau_e} = A \int_{0}^{\theta_{m}}d\theta (1-\cos\theta) {q} (N_q +1)\frac{1-f(\ve - \hbar \omega_q)}{1-f(\ve)},
\end{equation}
where
\begin{equation}
A = \frac{1}{2\pi}\frac{m}{\hbar^2}\frac{D^2}{\rho_0 v_{s}},
\end{equation}
and $\theta_{m}$ is determined by the following energy-momentum conservation laws for the scattering process $\vk \rightarrow \vk'$
\begin{eqnarray}
\ve' & = &\ve \pm \hbar \omega_q, \nonumber \\
q & = & \left [ k^2 + k'^2 -2 k k' \cos \theta \right]^{1/2}.
\end{eqnarray}

\begin{figure*}[tb]
\vspace{10pt}%
\includegraphics[width=1.\linewidth]{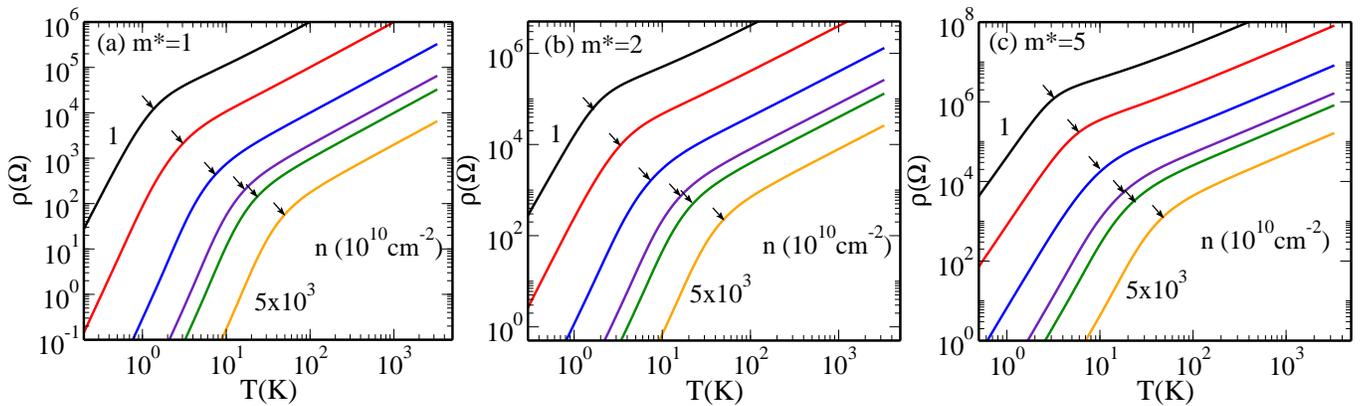}
\caption{
Calculated resistivity as a function of temperature for larger effective masses and for different densities ($n\times 10^{10}$cm$^{-2}$, where $n=1, \; 10, \; 10^2, \; 5\times 10^2, \; 10^3, \; 5\times10^3$ from top to bottom). The small arrows indicate the crossover temperatures, $T_c$.
}
\label{fig:high_m}
\end{figure*}

For $\hbar \omega_q \ll E_F$ the scattering of electrons by acoustic phonons can be considered quasi-eleastically. For quasi-elastic scattering, $k=k'$, and $q=2k \sin(\theta/2)$.
Changing the integration variable $\theta$ to $q$ and with $q_m = 2k$, 
we have
\begin{equation}
\frac{1}{\tau_a} = {4}A \int_{0}^{2k}\frac{dq}{\sqrt{1-(q/2k)^2}} \left ( \frac{q}{2k} \right )^3 N_q,
\label{eq:tau_a}
\end{equation}
\begin{equation}
\frac{1}{\tau_e} = 4A \int_{0}^{2k}\frac{dq}{\sqrt{1-(q/2k)^2}} \left ( \frac{q}{2k} \right )^3 (N_q +1).
\label{eq:tau_e}
\end{equation}
For $\hbar \omega_{\vq} \ll k_B T$, $N_q \sim N_q+1 \sim k_B T/\hbar \omega_q$. Then we have
\begin{equation}
\frac{1}{\tau(\ve)} =
\frac{m}{\hbar^3}\frac{D^2}{2\rho_0 v_{s}^2} k_BT.
\end{equation}
Since the scattering time is independent of the energy the resistivity is simply given by a linear-in-$T$ behavior \cite{KawamuraPRB1990}
\begin{equation}
\rho(T)  \propto T.
\end{equation}
In 2D this linear behavior holds even for a non-degenerate electron system, $T \gg T_F$.
However, in the BG regime, $T \ll T_{BG}$, the scattering rate is strongly reduced by the thermal occupation factors because the phonon population decreases exponentially. In the low-temperature limit (BG regime), the resistivity is given by $\rho \propto T^4$. 
The low-temperature Bloch-Gr\"{u}neisen regime ($T<T_{BG})$, where the phonon scattering effect on the resistivity is suppressed strongly, is mostly dominated by impurity scattering in electronic materials although in very clean systems, electron-electron scattering could also contribute to the low temperature resistivity.  Our main focus is the high-temperature linear-in-$T$ resistivity and the density-dependent crossover temperature, $T_c (n)$, above which this linearity applies.

\begin{figure}[b]
\vspace{10pt}%
\includegraphics[width=.8\linewidth]{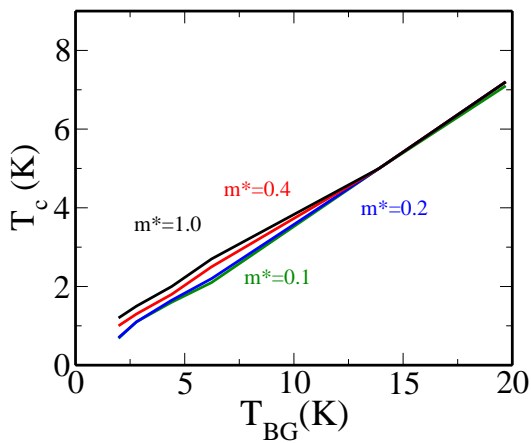}
\caption{
The crossover temperatures ($T_c$) obtained in Fig.~1 is shown as a function of $T_{BG}$. The relation between them is approximately is given by $T_c \approx 0.35 \; T_{BG}$ regardless of the effective mass.
}
\label{fig:cross}
\end{figure}

Throughout this paper we use the parameters\cite{price} corresponding to GaAs except effective mass $m$ (which we vary): the sound velocity $v_s = 5.14 \times 10^5$ cm/s, the deformation potential $D=12$ eV, and the 3D mass density $\rho_{3D} = 5.32$ g/cm$^3$. The characteristic width of 2D system $w=10$ nm is used. 
We note that most of these parameters, except for the effective mass, do not vary by more than a factor of $2-3$ in most electronic materials including even the materials often touted as strange metals.
We note that in GaAs (as in most semiconductors) the electron-acoustic phonon interaction strength is small, with $\lambda \ll 1$, making the weak-coupling approximation well-valid.

In Figs.~\ref{fig:rho_2d_m1} and \ref{fig:high_m} we show the calculated 2D resistivity limited by the acoustic phonon scattering as a function of temperature for different effective masses and electron densities. 
The calculated resistivity increases with high power ($T^4$) at low temperatures ($T<T_{BG}$) and linearly at high temperatures ($T>T_{BG}$). The calculated 2D resistivity at high temperatures increases linearly even in the non-degenerate temperature region ($T \gg T_F$).
In Figs.~\ref{fig:rho_2d_m1} and \ref{fig:high_m}  the small arrows indicate the crossover temperatures ($T_c$), at which the resistivity changes the power law from high ($T^4$) to linear ($T$) behavior as temperature increases. 
We find that the relation between the crossover temperature and $T_{BG}$ is approximately given by 
$T_c \approx 0.35 \; T_{BG}$ regardless of the effective mass as shown in Fig.~\ref{fig:cross}.
For 2D systems, $T_{BG} =2\hbar v_s k_F \propto n^{1/2}$, which is independent of the effective mass. Thus, we expect $T_c \propto n^{1/2}$ regardless of $m$. However, the ratio of $T_{BG}$ to $T_F$ is given by $T_{BG}/T_F = 7.0 m^*/\sqrt{\tilde{n}}$, where $m^*=m/m_e$ with $m_e$ being electron bare mass and  $\tilde{n}$ is the density measured in unit of $10^{10}$ cm$^{-2}$. At high effective masses and low densities 
we have $T_{BG}/T_F \gg 1$. 
(Note that this limit of $T_{BG}  \gg T_F$ is never achieved in any regular metals where $T_F \gg T_{BG}>T_D$.)
In this limit, the empirical relation between $T_c$ and $T_{BG}$ does not hold, i.e., $T_c \approx 0.35 \; T_{BG}$ relation deviates for high effective masses and low carrier densities.  
But the approximate relation $T_c  \sim 0.35$ $T_{BG}$ holds well as long as $T_{BG}<T_F$.  Below we consider the situation (which applies in high-density system such as normal metals) where $T_{BG}>T_D$, where $T_D$ is the standard Debye temperature for the acoustic phonons.  For $T_D < T_{BG}$, the phonon energy cut off is $T_D$ which must explicitly be taken into account.

\begin{figure}[tb]
\vspace{10pt}%
\includegraphics[width=1.\linewidth]{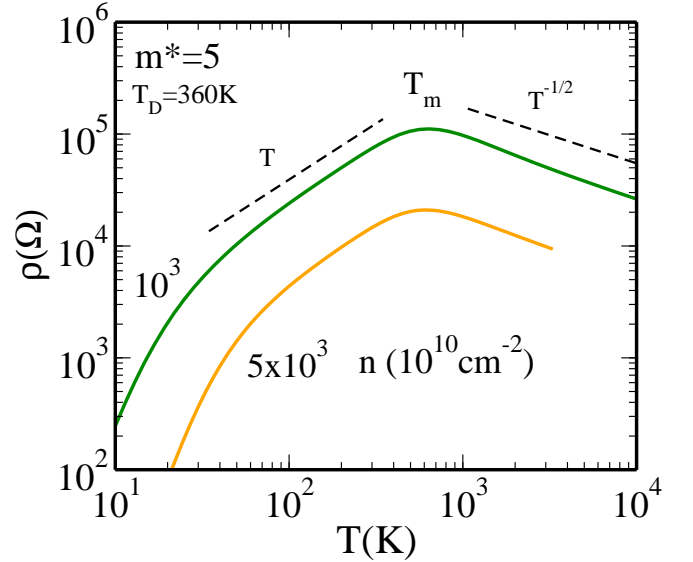}
\caption{
Calculated resistivity as a function of temperature with GaAs parameters and the effective mass $m^*=m/m_e=5$. Here we have $\rho \sim T$ for $T<T_0$ and $\rho \sim T^{-1/2}$ for $T>T_0$. The high temperature cross over (i.e., maximum resistivity) occurs  at $T_m \sim T_0/3$. As shown in Eq.~(19) we have  $T_0 \gg  T_D$, the Debye temperature.
}
\label{fig:high_debye}
\end{figure}

We now consider the Debye model of the acoustic phonon, where the total number of acoustic phonon modes is finite and the phonon frequencies must be smaller than the highest frequency $\omega_D$, the Debye frequency. Since the allowed phonon wave vectors are limited by $q \le q_D = \omega_D/v_{s}$,  the upper bound of the integrals in Eqs.~(\ref{eq:tau_a}) and (\ref{eq:tau_e}) is given by $q_{m}=$ min$(q_D, 2k)$.
If the electron energy in Eq.~(\ref{eq:tau_2d}) is smaller than $\ve_c$ corresponding to the wave vector $2k=q_D$, i.e.,
\begin{equation}
\ve < \ve_c = \frac{\hbar^2}{8m}\frac{\omega_D^2}{v_{s}^2},
\end{equation}
then all phonon wave vectors in Eqs.~(\ref{eq:tau_a}) and (\ref{eq:tau_e}) must be smaller than $q_D$, i.e., $q_{m}=2k$.
The temperature corresponding to $\ve_c$ becomes
\begin{equation}
T_0 = \ve_c/k_B = \frac{1}{8m} \frac{k_BT_D^2}{v_{s}^2},
\end{equation}
where $T_D$ is the Debye temperature. 
This $T_0$ is a new energy scale for the coupled electron-phonon system where the phonon phase space restriction arising from the Debye cut off becomes relevant quantitatively.
For GaAs $T_D=360$K, and we find $T_0 = 9360/m$ K, where $m$ is the effective mass of the system. Note that $T_0$ depends explicitly on the effective mass and is unphysically large unless the effective mass is large ($m^* \gg 1$).
When the electron energy in Eq.~(\ref{eq:tau_2d}) is greater than $\ve_c$, $q_m = q_D$ and the upper limit of Eqs.~(\ref{eq:tau_a}) and (\ref{eq:tau_e}) is bounded by $q_D$. Due to this restriction of the scattering angle, the scattering rate and resistivity start decreasing at high temperatures  $T>T_0=9360/m$ K.  The actual decrease begins at $T_m=T_0/3$.  This phonon phase space restricted (very) high temperature decrease of resistivity has not been considered before in the literature.  This high-temperature decrease of resistivity looks similar to the so-called resistivity saturation phenomenon, but seems to happen here at higher temperatures.
For example, in GaAs, with $m=0.07m_e$, $T_0 \sim 100,000$K, which is obviously a temperature scale of no physical relevance.  For a system with a much larger effective mass, however, it is , in principle, possible for this phonon phase space restriction physics to manifest itself at a very high temperature scale.  We give an example in Fig.~\ref{fig:high_debye} where $T_m \sim 700$K for a system with a large effective mass of $m^*=m/m_e = 5$.
(We note as an aside that most materials manifesting resistivity saturation phenomenon typically has an effective mass larger than $m_e$.)

To explicitly see the temperature dependence of the scattering time with the Debye cutoff effect for $T\gg T_F$, we rewrite Eq.~(\ref{eq:tau_2d}) for the non-degenerate Fermi distribution function, i.e., $f(\vek) \sim e^{-\frac{\vek}{k_BT}}$.
\begin{equation}
\langle \tau \rangle = \frac{\int d \ve \; \tau \ve e^{-\ve/k_BT}}{\int d\ve \; \ve e^{-\ve/k_BT}}.
\label{eq:high_sigma}
\end{equation}
We rewrite the numerator of Eq.~(\ref{eq:high_sigma}) as
\begin{equation}
I = \left ( \int_0^{\ve_c} +  \int_{\ve_c}^{\infty} \right )d \ve \; \tau \ve e^{-\ve/k_BT} = I_{l} + I_{h},
\end{equation}
where $I_l$ is the low energy contribution
and $I_h$ is the high energy contribution.
Since $2k<q_D$ in $I_l$, we have 
\begin{equation}
I_l \propto T \int_0^{x_c} dx \; x e^{-x} = 1-(1+x_c)e^{-x_c},  
\end{equation}
where $x_c = T_0/T$. Thus, we have
\begin{eqnarray}
I_l & \propto& T \;\; {\rm for} \;\; x_c \gg 1 \; {\rm or} \; T \ll T_0 \nonumber \\
     & \propto& T x_c^2 \sim \frac{1}{T} \;\; {\rm for } \;\; x_c \ll 1 \; {\rm or} \;T \gg T_0.
\end{eqnarray}     
In $I_h$, $\ve > \ve_c$, and we have $q_{m} = q_D$ in Eqs.~(\ref{eq:tau_a}) and (\ref{eq:tau_e}). Then,
the scattering time becomes
\begin{equation}
\frac{1}{\tau} \propto \left(\frac{q_D}{2k} \right )^3 T,
\end{equation}
i.e., $\tau(\ve) \sim \ve^{3/2} T^{-1}$.
By using the asymptotic behavior of $I_h$ we obtain
\begin{eqnarray}
I_h & \propto& T^{5/2}x_c^{5/2}e^{-x_c} \sim e^{-T_c/T} \;\; {\rm for} \;\; x_c \gg 1 \; {\rm or} \; T \ll T_0 \nonumber \\
     & \propto& T^{5/2}  \;\; {\rm for } \;\; x_c \ll 1 \; {\rm or} \;T \gg T_0.
\end{eqnarray}     
Thus, we find that for $T<T_0$, $I_l$ dominates over $I_h$, and for $T>T_0$ $I_h$ dominates over $I_l$. 
Since the denominator of Eq. (21) gives $T^2$ contribution we finally have 
\begin{eqnarray}
\langle \tau \rangle & \propto& T^{-1} \;\; {\rm for} \; \; T \ll T_0 \nonumber \\
     & \propto& T^{1/2}  \;\; {\rm for } \; \;T \gg T_0,
\end{eqnarray}     
and the resistivity becomes
\begin{eqnarray}
\rho(T) & \propto& T \;\; {\rm for} \; \; T < T_0 \nonumber \\
     & \propto& T^{-1/2}  \;\; {\rm for } \; \;T \gg T_0.
\label{eq:rho_td}
\end{eqnarray}     

\begin{figure*}[htb]
\vspace{10pt}%
\includegraphics[width=.75\linewidth]{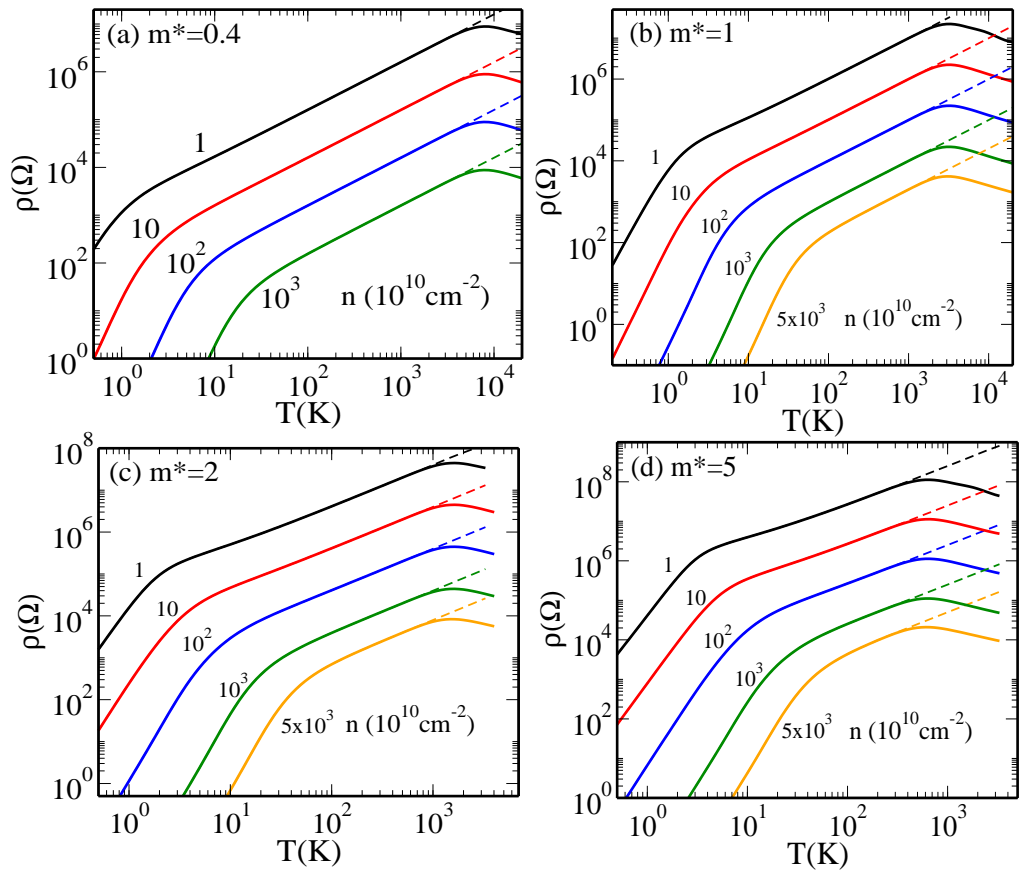}
\caption{
Resistivity $\rho$ vs. temperature for different effective masses and for carrier densities shown in the figures with momentum cutoff corresponding to the maximum Debye phonon modes. The cutoff wave vector is given by $q_D = k_B T_{D}/\hbar v_{s}$, where $T_{D}$ is the Debye temperature. 
}
\label{fig:high_debye1}
\end{figure*}

\begin{figure*}[htb]
\vspace{10pt}%
\includegraphics[width=.75\linewidth]{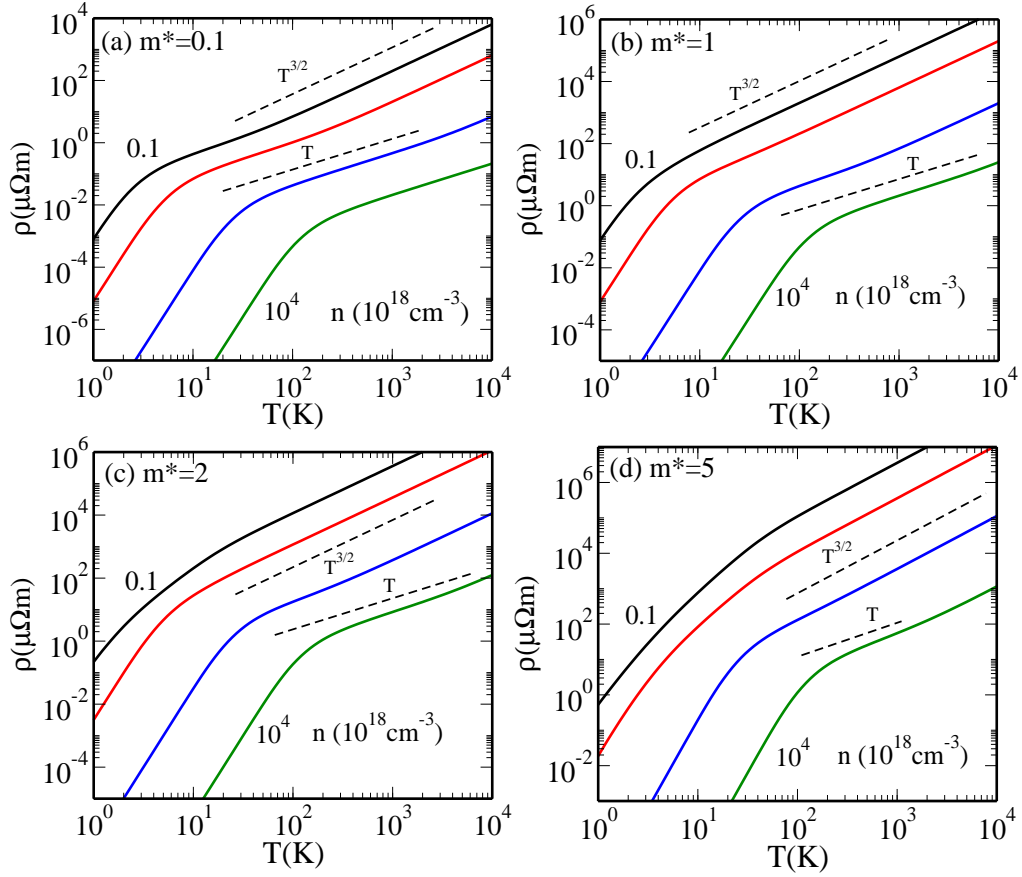}
\caption{
Calculated 3D resistivity as a function of temperature for different effective masses and densities ($n\times 10^{18}$cm$^{-3}$, where $n=0.1, \; 1, \; 10^2, \; 10^4$ from top to bottom) without Debye momentum cutoff. The parameters corresponding to GaAs are used in this figure. 
At low temperatures $T < T_{BG}$ $\rho \propto T^5$. At high temperatures we have a linear temperature dependence of resistivity for highly degenerated systems ($T/T_F \ll1$), but for nondegenerated systems the resistivity increases as $T^{3/2}$.
}
\label{fig3d}
\end{figure*}

The numerically calculated temperature dependent resistivity, which agrees with our asymptotic analytical formula, 
is shown in Figs.~\ref{fig:high_debye} and \ref{fig:high_debye1}. 
The Debye cutoff has no effect on the resistivity at low temperatures, but at high temperatures 
the resistivity is suppressed due to the restriction of the scattering angle. Based on our calculation we have $m T_m \approx$ constant, where $T_m$ ($\sim T_0$) is the temperature at the maximum resistivity (which is found empirically through the numerical calculation). 
Figure \ref{fig:high_debye} shows the temperature dependent resistivity with GaAs parameters and the effective mass $m = 5 m_e$ for two different densities. As shown in Fig.~\ref{fig:high_debye}  the resistivity is divided into three distinct regions, BG region at low temperatures with $\rho \propto T^4$, linear region 
at intermediate temperatures
with $\rho \propto T$, and scattering phase space limited region with Debye cut off at high temperatures with $\rho \propto T^{-1/2}$.
The numerically calculated high temperature crossover of the resistivity from $T$ to $T^{-1/2}$ occurs at $T_m \sim T_0/3$, which is independent of the electron density. Note that $T_0$ is a function of effective mass only.

In Fig.~\ref{fig:high_debye1} we show the calculated 2D resistivity as a function of temperature for different masses. Even though the low temperature crossover ($T_c$) from $T^4$ to $T$ behavior varies with the carrier density (since $T_{BG} \sim  k_F$),
the high temperature crossover ($T_m$) does not depend on the density for a fixed effective mass, which can be understood from the density dependence of $T_F$, $T_{BG}$  and $T_D$. $T_m$ is approximately given by $T_m \sim T_0/3 = k_B T_D^2/(24 v_c^2 m) \approx 3120/m^*$ K for 2D GaAs systems. Thus, $T_m$ decreases with increasing effective mass, but it is independent of the electron density. 
The low temperature crossover $T_c$ increases with density, i.e., $T_c \approx 0.35 T_{BG} \approx 0.7 \sqrt{\tilde{n}}$, where $\tilde{n}$ is measured in unit of $10^{10}$ cm$^{-2}$, but it is independent of the effective mass. Thus, at low electron densities and for a low effective mass system, the linearly increasing resistivity can manifest in a wide range of temperatures
with the resistivity being linear down to very low temperatures in dilute 2D systems.
As shown in Fig.~\ref{fig:high_debye1}(b) where $m^*=5$, the linear region is limited in the high density limit
since $T_c$ ($T_m$) is high (low) in the large density (mass) situation.
We emphasize that our finding of an effective resistivity saturation type phenomenon (actually a decreasing resistivity with increasing $T$) at very high temperatures ($T>T_m \sim T_0/3$) may be physically relevant only for very high carrier effective masses with $m^*>5$.  For smaller effective masses (with $m^* \sim 1$ or below), $T_0 >10,000$K is unphysically high without any experimental relevance.


\begin{figure}[htb]
\vspace{10pt}%
\includegraphics[width=1.\linewidth]{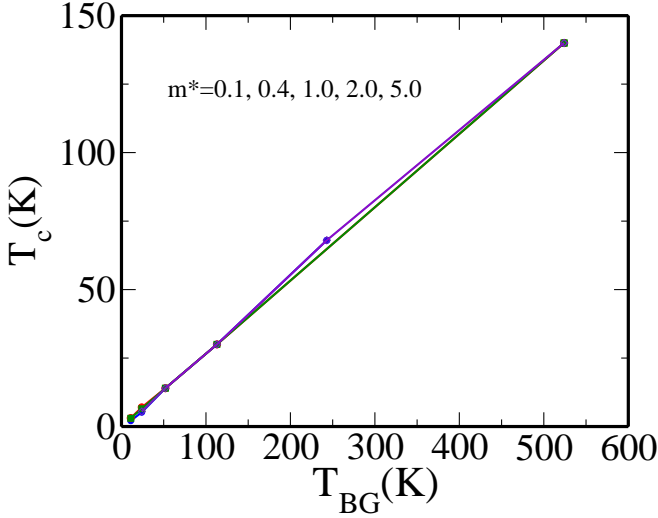}
\caption{
The crossover temperature, $T_{c}$, from BG to linear temperature dependence in 3D systems is shown as a function of $T_{BG}$, where $T_{c}$ is obtained from Fig. 6. The relation between $T_c$ and $T_{BG}$ in 3D  is given by $T_c \approx 0.27 T_{BG}$.
}
\label{fig:ratio3D}
\end{figure}

\section{Resistivity in 3D systems}

The 3D density of states is given by $N(\ve) = \sqrt{2\ve}m^{3/2}/(\pi^2\hbar^3)$. Using the 3D density of states the conductivity can be expressed as $\sigma = ne^2 \langle \tau \rangle/m$, where $n$ is the 3D electron density and the energy averaged scattering time $\langle \tau \rangle$ is give by \cite{Ziman2}
\begin{equation}
\langle \tau \rangle = \frac{\int d \ve \; \tau(\ve) \ve^{3/2} \left( - \frac{\partial f(\ve)}{\partial \ve} \right ) }{\int d\ve \; \ve^{3/2} \left( - \frac{\partial f(\ve)}{\partial \ve} \right ) }.
\label{tau3d0}
\end{equation}
The energy dependent 3D scattering time given in Eq.~(\ref{tau_3d})  with 3D wave vectors ($\vq$, $\vk$, and $\vk'$)  becomes in the quasi-elastic limit, $\hbar \omega_q \ll k_BT$
\begin{eqnarray}
\frac{1}{\tau(\ve)} & = & \frac{1}{4\pi}\frac{D^2}{2\rho_{3D} v_{s}} \left ( \frac{2m}{\hbar^2} \right )^{3/2} \sqrt{\ve} \nonumber \\
& & \times\int_0^{\pi} \sin \theta d\theta (1-\cos \theta) q (2N_q +1).
\end{eqnarray}
With $q = 2k\sin (\theta/2)$ and $N_q \approx k_BT/\hbar \omega_q$ we have
\begin{eqnarray}
\frac{1}{\tau(\ve)} &= &\frac{4}{2\pi \hbar} \frac{D^2}{\rho_{3D} v_{s}^2}  \left ( \frac{2m}{\hbar^2} \right )^{3/2} (k_BT) \sqrt{\ve} \nonumber \\
& & \times \int_0^{2k} \frac{dq}{2k} \left ( \frac{q}{2k} \right )^3.
\end{eqnarray}
The upper limit of integral $2k$ must be changed to $q_D$ if $q_D < 2k$, i.e., 
$q_{m}=$ min$(q_D, 2k)$, where $q_D = \omega_D/v_{s}$. 
For $q_D > 2k$ we have
\begin{eqnarray}
\frac{1}{\tau(\ve)} & = & \frac{1}{2\pi \hbar} \frac{D^2}{\rho_{3D} v_{s}^2}  \left ( \frac{2m}{\hbar^2} \right )^{3/2} (k_BT) \sqrt{\ve} \nonumber \\
& \propto & T \sqrt{\ve},
\label{eq:tau_3D1}
\end{eqnarray}
and for $q_D < 2k$
\begin{eqnarray}
\frac{1}{\tau(\ve)} & = &\frac{1}{2\pi \hbar} \frac{D^2}{\rho_{3D} v_{s}^2}  \left ( \frac{2m}{\hbar^2} \right )^{3/2} (k_BT) \sqrt{\ve} \left ( \frac{q_D}{2k} \right )^4, \nonumber \\
& \propto & T \ve^{-3/2}.
\label{eq:tau_3D2}
\end{eqnarray}
Using Eq.~(\ref{eq:tau_3D1}) we find that the resistivity increases linearly with increasing temperatures (i.e., $\rho \propto T$ for $T>T_{BG}$) for a degenerate system ($T\ll T_F$).
For the non-degenerate case ($T\gg T_F$) 
the energy averaged 3D scattering time becomes
\begin{equation}
\langle \tau \rangle = \frac{\int d \ve \; \tau(\ve) \ve^{3/2} e^{-\ve/k_BT}}{\int d\ve \; \ve^{3/2} e^{-\ve/k_BT}},
\label{tau3d}
\end{equation}
and in the absence of any momentum cutoff, the energy dependent scattering time is given by $\tau(\ve) \sim T^{-1} \ve^{-1/2}$. Thus, the resistivity increases as $T^{3/2}$ for $T\gg T_F$. Unlike the 2D system where the resistivity increases linearly for all temperatures until a saturation sets in at $T_m$, the phonon limited 3D resistivity goes as
\begin{eqnarray}
\rho(T) &\sim& T \;\; {\rm for} \;\; T_{BG}<T \ll T_F  \nonumber \\
\rho(T) &\sim& T^{3/2} \;\; {\rm for} \;\;  T \gg T_F. 
\end{eqnarray}

In Fig.~\ref{fig3d} we show the calculated 3D resistivity as a function of temperature for different effective masses and electron densities. We use the parameters corresponding to GaAs except that effective mass is a variable parameter. 
At low temperatures $T < T_{BG}$ the resistivity decreases as $T^5$ with decreasing temperature due to the exponential decrease of phonon population in the Bloch-Gr\"{u}neisen regime. As temperature increases
the power law changes from $T^5$ to linear ($T$) as long as $T_{BG} < T_F$. However, for $T_{BG} > T_F$ we find a $T^5$ to $T^{3/2}$ transition 
without the resistivity manifesting any linear-$T$ behavior.  In normal metals, where the linear-in-$T$ resistivity is routinely observed at room temperatures, $T_F \gg T_{BG}$ (or $T_D$), and hence the $T^{3/2}$ high-temperature behavior cannot manifest itself.
Unlike 2D systems, a linear temperature dependence of resistivity at high temperatures in 3D applies only to degenerate systems ($T/T_F \ll1$). For 3D non-degenerate systems, the resistivity increases as $T^{3/2}$ and the linear temperature dependent region is not apparent.
For systems with low Fermi temperatures ($T_F \sim T_{BG}$) the linear region is very narrow as the resistivity does not show the linear behavior for $T_{BG} > T_F$, which is the situation for high effective mass and low electron densities. As shown in Fig.~\ref{fig3d}(a) the transition from low-$T$ to intermediate temperature $T$ region with increasing temperature is clearly seen for high densities and low effective masses which correspond to the $T_{BG} \ll T_F$.

As in the 2D system the crossover temperature $T_c$ from a high power law to a linear region in the 3D case is closely related to $T_{BG}$. 
The relation between $T_c$ and $T_{BG}$ is approximately given by 
$T_c \approx 0.27 \; T_{BG}$ in 3D systems regardless of the effective mass as shown in Fig.~\ref{fig:ratio3D}. 
For 3D systems, $T_{BG} =2\hbar v_s k_F \propto n^{1/3}$, which is independent of the effective mass. Thus, we expect $T_c \propto n^{1/3}$ regardless of $m$. 

\begin{figure*}[htb]
\vspace{10pt}%
\includegraphics[width=.80\linewidth]{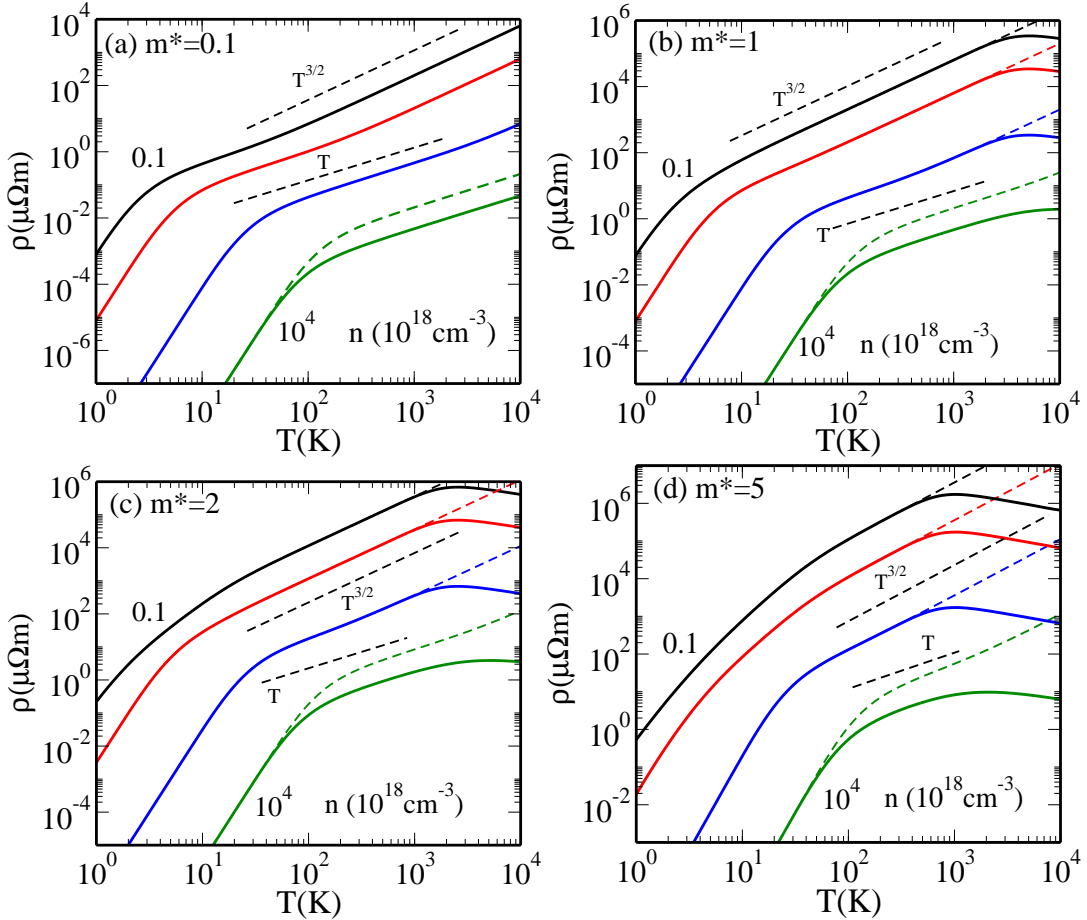}
\caption{
Calculated 3D resistivity as a function of temperature for different effective masses and densities ($n\times 10^{18}$cm$^{-3}$, where $n=0.1, \; 1, \; 10^2, \; 10^4$ from top to bottom) with Debye momentum cutoff. The solid lines represent the results with Debye cutoff. The large suppression of the scattering rate at high densities arises from the restriction of the scattering angle, especially for $T_F > T_0$.}
\label{fig3dr}
\end{figure*}

\begin{figure*}[htb]
\vspace{10pt}%
\includegraphics[width=.90\linewidth]{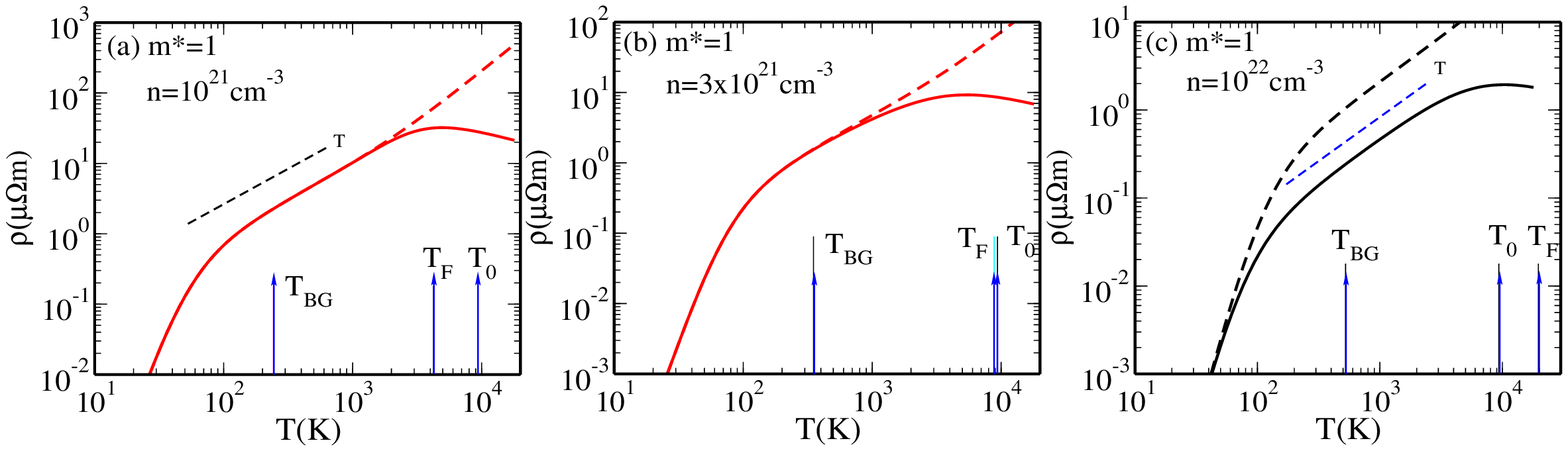}
\vspace{10pt}%
\includegraphics[width=.90\linewidth]{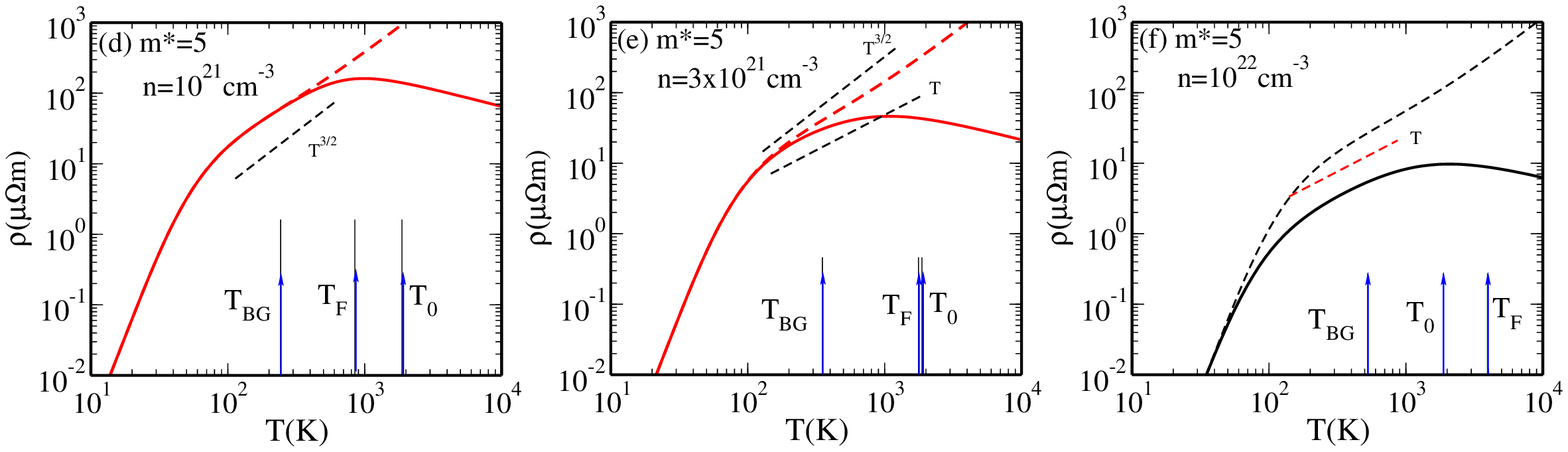}
\caption{
(a) (b), and (c) show the calculated 3D resistivity as a function of temperature for  $m^*=1$ and for  $T_F < T_0$, $T_F \approx T_0$, and $T_F > T_0$, respectively.
For $T_0 < T_F$, the scattering angle is restricted even for the degenerated case, $T_F \gg T$, which gives rise to the large suppression of the resistivity.
(d) (e), and (f) show the resistivity  for $m^*=5$  and for  $T_F < T_0$, $T_F \approx T_0$, and $T_F > T_0$, respectively. The dashed (solid) lines indicated the calculated resistivity without (with) momentum cut off corresponding to the Debye frequency. 
}
\label{compare3d5}
\end{figure*}

To obtain the high temperature behavior in the presence of the Debye momentum cutoff we rewrite the numerator of Eq.~(\ref{tau3d}) as 
\begin{equation}
I = \left ( \int_0^{\ve_c} + \int_{\ve_c}^{\infty} \right ) d \ve \; \tau \ve^{3/2} e^{-\ve/k_BT} = I_{l} + I_{h},
\end{equation}
where
$\ve_c = \frac{\hbar^2}{8m}\frac{\omega_D^2}{v_{s}^2}$.  We also denote $T_0=\ve_c/k_B$.
$I_l$ restricts $2k<q_D$, and the scattering time becomes $\tau  \sim T^{-1}\ve^{-1/2}$.
Then, we get
\begin{eqnarray}
I_l & \propto&      T  \;\; {\rm for } \;\; x_c \gg 1 
\nonumber \\
     & \propto&  T x_c^2 \sim \frac{1}{T} \;\; {\rm for} \;\; x_c \ll 1,
\end{eqnarray}     
where $x_c = \ve_c/k_BT$. Thus, the leading order behavior for $I_l$ becomes
\begin{eqnarray}
I_l & \propto&      T \;\; {\rm for } \;\; T \ll T_0  
\nonumber \\
     & \propto&  T^{-1}  \;\; {\rm for} \;\; T \gg T_0.
\end{eqnarray}     
In $I_h$ the energy is always greater than $\ve_c$, $\ve > \ve_c$, then we have $q_{m} = q_D$. From Eq.~(\ref{eq:tau_3D2})
\begin{equation}
{\tau} \propto T^{-1} \ve^{3/2}
\end{equation}
and we have
\begin{eqnarray}
I_h & \propto& T^{3}e^{-x_c}x_c^3   \;\; {\rm for } \;\; x_c \gg 1  \nonumber \\
     & \propto& T^{3} \left [ 6 - x_c^4/4 \right ]  \;\; {\rm for } \;\; x_c \ll 1
\end{eqnarray}     
and the leading order behavior becomes
\begin{eqnarray}
I_h & \propto& e^{-T_0/T}   \;\; {\rm for } \;\; T \ll T_0  \nonumber \\
     & \propto& T^{3}  \;\; {\rm for } \;\; T \gg T_0
\end{eqnarray}     
Thus, for $T<T_0$, $I_l$ dominates over $I_h$, and for $T>T_0$ $I_h$ dominates over $I_l$. 
Since the denominator of Eq.~(\ref{tau3d}) gives $T^{5/2}$ we have finally 
\begin{eqnarray}
\langle \tau \rangle & \propto& T^{-3/2} \;\; {\rm for} \; \; T \ll T_0 \nonumber \\
     & \propto& T^{1/2}  \;\; {\rm for } \; \;T \gg T_0,
\end{eqnarray}     
and the resistivity becomes
\begin{eqnarray}
\rho(T) & \propto& T^{3/2} \;\; {\rm for} \; \; T \ll T_0 \nonumber \\
     & \propto& T^{-1/2}  \;\; {\rm for } \; \;T \gg T_0.
\end{eqnarray}     
We emphasize these results only apply to the non-degenerate system. For degenerate systems we simply have  $\rho \sim T$ as long as  $T_{BG}<T<T_F$.  Thus, the temperature dependence depends crucially on whether $T_F$ is larger or smaller than $T_0$.  In 3D normal metals, $T_F \gg T_0$.  In dilute exotic electronic materials (with low carrier density, large carrier effective mass, and/or large Debye energy), $T_F<T_0$ may be achieved.

In Fig.~\ref{fig3dr} we show the calculated resistivity of a 3D system with the Debye momentum cutoff ($q_D$) corresponding to the Debye frequency. The GaAs parameters are used as in the 2D case. 
The calculated resistivity at low temperatures is not affected by the cutoff (since $T_{BG}<T_D$ throughout or our parameters), but at high temperatures the resistivity is suppressed due to the restriction of the scattering angle, decreasing as $T^{-1/2}$ with increasing temperature. Thus, the resistivity has a maximum at $T_m$ and the calculated results show that the product of effective mass and $T_m$ is independent of the density, i.e., $m T_m \sim$ constant.
As shown in Fig.~\ref{fig3dr}, at low temperatures $\rho \propto T^5$ and at high temperatures ($T>T_0$) $\rho \propto T^{-1/2}$. In the intermediate region ($T_{BG}<T<T_0$) the resistivity increases linearly or as $T^{3/2}$, depending on whether $T_{BG} < T_F$ or $T_{BG} > T_F$, respectively. 
The numerically calculated high temperature crossover occurs at $T_m \sim T_0/2.5$, which is independent of the electron density. Note that $T_0$ is a function of carrier effective mass only and not of the carrier density.
The wide temperature range of linearly increasing 3D resistivity is achieved only when the conditions of $T_{BG} \ll T_F \ll T_0 $ is satisfied. 
This inequality is satisfied by all 3D metals, where the linear-in-$T$ resistivity typically persists over a large temperature regime (50 -- 1000K).

Fig.~\ref{compare3d5} shows how the linearly temperature dependent region of resistivity can be modified by
the interplay among $T_{BG}$, $T_F$, and $T_0$. In Fig.~\ref{compare3d5}
the 3D resistivity as a function of temperature is shown by changing the order between $T_0$ and $T_F$. To obtain the linear $T$ region we set $T_{BG}<T_F$ and $T_0$.
Figs.~\ref{compare3d5}(a) (b), and (c) show the results for  $T_F < T_0$, $T_F \approx T_0$, and $T_F > T_0$, respectively for the effective mass $m=m_e$.
Figs.~\ref{compare3d5}(d) (e), and (f) show the results for  $T_F < T_0$, $T_F \approx T_0$, and $T_F > T_0$, respectively for the effective mass $m=5m_e$.
For $T_0 < T_F$, the scattering angle is restricted even for the degenerate case, $T_F \gg T$, which gives rise to the large suppression of the resistivity.
We show the linearly increasing resistivity between $T_c$ and $T_m$. Note that the low temperature crossover is $T_c \approx 0.27 T_{BG}$ and the high temperature crossover is $T_m \approx 0.4 T_0$.
Thus, a large separation between $T_{BG}$ and $T_0$ and the condition of $T_{BG}<T_F<T_0$ are both required to get a large window of linearly increasing resistivity with temperature in 3D systems. 
We note that, similar to the 2D situation, the actual magnitude of $T_0$ in 3D is unphysically large unless the effective mass is large -- e.g., $T_0 \sim 10^4$ ($10^3$) K in Figs.~\ref{fig3dr} and  \ref{compare3d5} for $m^*=1$ (5).  Thus, the phonon phase space restriction induced high-temperature resistivity decrease may only be relevant for systems with large effective mass (as well as low carrier density), and is not physically relevant for ordinary 3D metals.

\begin{figure*}[t]
\vspace{10pt}%
\includegraphics[width=1.0\linewidth]{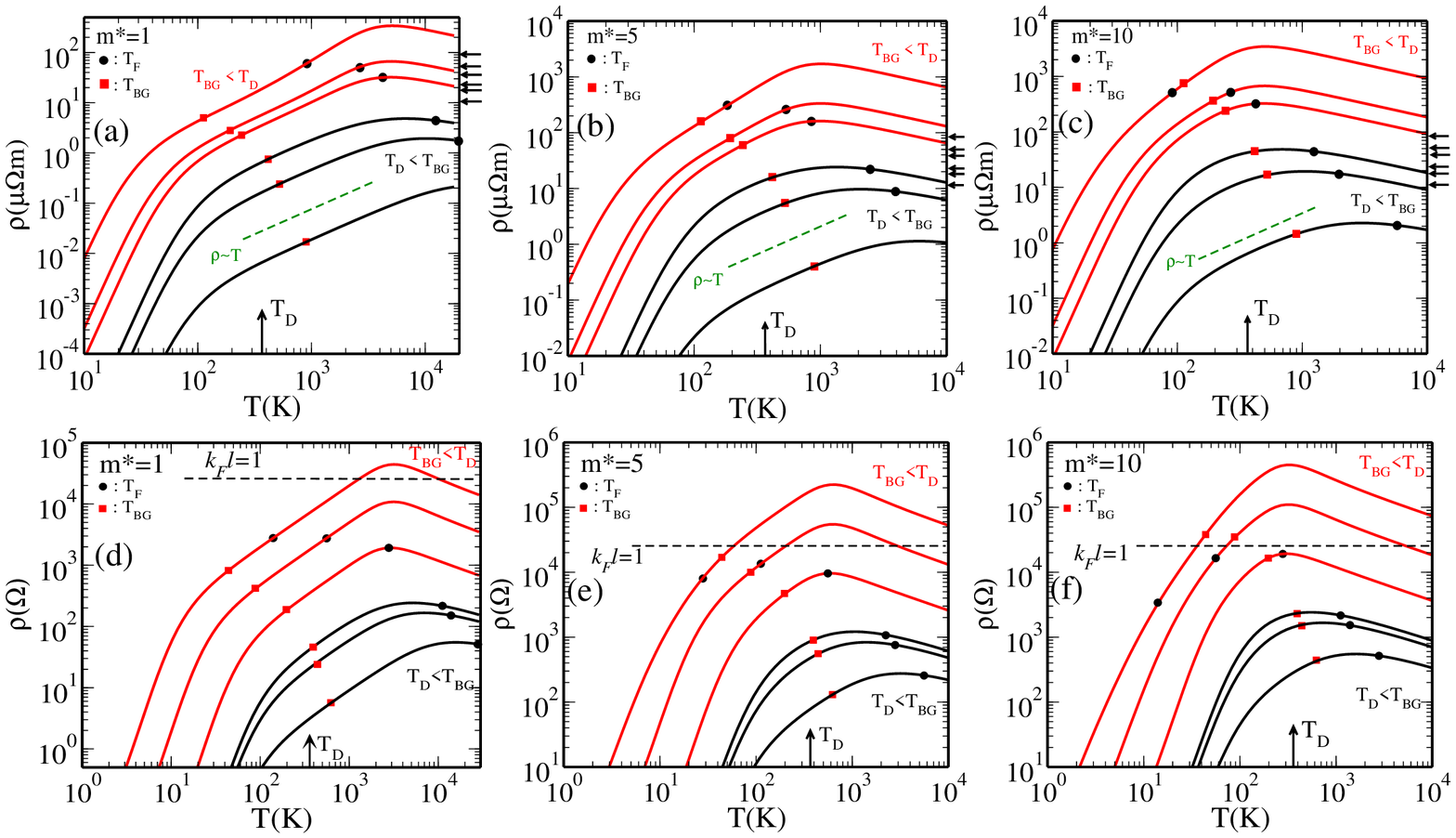}
\caption{
(a), (b), and (c) show the calculated 3D resistivity with $m^* =1$, 5, and 10, respectively, for various densities, $n=10^{20}, \; 5\times10^{20}, \; 10^{21}, \; 5\times 10^{21}, \;10^{22}, \;5\times 10^{22}$ cm$^{-3}$ (top to bottom), which correspond to  $T_{BG}= $113, 193,  243, 416, 524 K for all masses and $T_F= 914/m^*, \;2671/m^*, \;4241/m^*, \;12401/m^*, \;19686/m^*, \;57564/m^*$ K. Black dots (red squares) indicate $T_F$ ($T_{BG}$). 
The temperature corresponding momentum cut off $T_0=9360$K and the Debye temperature $T_D=360K$. The small arrows on the right vertical axis indicate $k_F l = 1$ corresponding to the densities used in the resistivity results (top to bottom).  
The red (black) lines represent $T_{BG} < T_D$ ($T_{BG} > T_D$).
(d), (e), and (f) show the calculated 2D resistivity with $m^* =1$, 5, and 10, respectively, for various densities, $n=5\times10^{12}, \; 3\times10^{13}, \; 10^{14}, \; 4\times 10^{14}, \;5\times10^{14}, \; 10^{15}$ cm$^{-2}$ (top to bottom), which correspond to  $T_{BG}= $44, 88,  197, 394, 441, 623 K for all masses and $T_F= 140/m^*, \;560/m^*, \;2790/m^*, \;11165/m^*, \;13960/m^*, \;27915/m^*$ K. Black dots (red squares) indicate $T_F$ ($T_{BG}$). The dashed horizontal lines indicate $k_Fl = 1$ which is independent of the density in 2D. 
}
\label{fig:new}
\end{figure*}

In Fig~\ref{fig:new} we show the calculated 3D and 2D resistivity as a function of temperature for various densities with fixed effective mass $m^*=1$, 5, and 10. The Debye temperature and the temperature corresponding to the momentum cutoff are given by $T_D=360K$ and $T_0=9360$K, respectively, for both 2D and 3D. Depending on the carrier density and the effective mass the relative magnitudes 
among $T_{F}$, $T_{BG}$, and $T_D$ can be varied. The red lines indicate $T_{BG} < T_D$, the blue lines $T_D < T_{BG} $. For $T_{BG} < T_D$, the temperature corresponding to the maximum resistivity, $T_m$, is independent of  the density (or $T_{BG}$). However, for $T_{BG} > T_D$, $T_m$ increases with increasing density (or $T_{BG}$).
In Fig.~\ref{fig:new}(a), (b), (c) the small arrows on the right vertical axis indicate $k_F l = 1$ corresponding to the densities used in the results. In 3D $k_F l=1$ corresponds to $\rho = 3910/\tilde{n}^{1/3} \;\mu\Omega$m, where $\tilde{n}$ is the 3D density measured in units of $10^{15}$ cm$^{-3}$. However, in 2D, $k_Fl=1$ is independent of the density and it corresponds to $\rho = h/e^2$. The dashed horizontal lines in Fig.~\ref{fig:new}(d), (e), (f) indicate $k_Fl = 1$ (or $\rho = h/e^2$). 
Results of Fig.~\ref{fig:new} emphasize that the linearity-in-$T$ is a generic and physically relevant feature of the calculated resistivity, but the high-temperature resistivity saturation phenomenon is physically relevant only for rather high effective mass ($m=5m_e$) systems.  In addition, Fig.~\ref{fig:new} also demonstrates that any connection between the Mott-Ioffe-Reggel criterion\cite{ioffe,mott} and our theoretical resistivity decrease phenomenon is at best coincidental.

\begin{figure*}[t]
\vspace{10pt}%
\includegraphics[width=.90\linewidth]{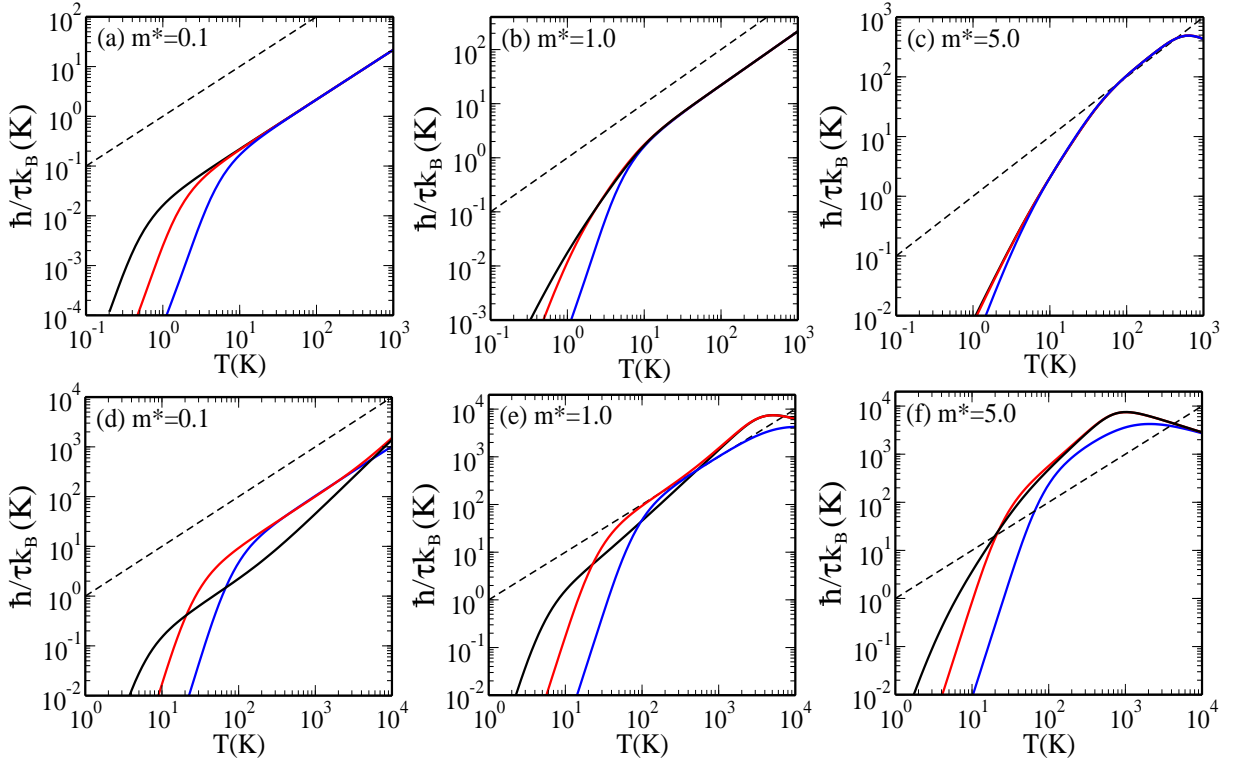}
\caption{
The calculated inverse scattering time ($1/\tau $) as a function of temperature for different masses and for different densities. The inverse scattering time is given in the unit of temperature, $\hbar/\tau k_B$. The Debye cutoff effects are included in the all calculation. The dashed line in all figures indicates $\hbar/\tau k_B=T$. (a) (b), and (c) show the 2D inverse scattering time for the effective mass $m^*=0.1,$ 1.0, 5.0, respectively. The black, red, blue lines are corresponding to the 2D electron density $n=10^{10}, \; 10^{11}, \; 10^{12}$ cm$^{-2}$ , respectively.
(d) (e), and (f) show the 3D inverse scattering time for $m^*=0.1,$ 1.0, 5.0, respectively. The black, red, blue lines are corresponding to the 3D electron density $n=10^{18}, \; 10^{20}, \; 10^{22}$ cm$^{-3}$ , respectively.
The same GaAs parameters are used for both 2D and 3D systems.
}
\label{fig:tau}
\end{figure*}

Finally in Fig.~\ref{fig:tau} we compare (for both 2D and 3D systems) the calculated inverse scattering time ($1/\tau$) as a function of temperature for different masses ($m=0.1, \;1.0, \; 5.0m_e$) and for different densities. 
The inverse scattering time is given in the unit of temperature, i.e., $\hbar/\tau k_B$.
For 2D systems the electron density $n=10^{10}, \; 10^{11}, \; 10^{12}$ cm$^{-2}$ are used, and for 3D system $n=10^{18}, \; 10^{20}, \; 10^{22}$ cm$^{-3}$ are used.
The Debye cutoff effects are included in the figures  and the same GaAs parameters are used for both 2D and 3D systems. 
The comparison between $1/\tau$ and $T$ as shown in Fig.~\ref{fig:tau} has implications for the currently active debate on the role of a possible scale-invariant Planckian scattering rate in the transport properties of putative strange metals.  
We note that the scattering rate could be smaller or larger than $k_BT$ depending entirely on system parameters and is a completely nonuniversal quantity with no particular significance whatsoever with respect to its magnitude relative to temperature.
We comment more on this topic in Sec. V below.

\section{Transport in Dirac Materials}

We briefly consider phonon scattering limited transport in Dirac materials at high temperatures ($>T_{BG}$) where a linear-in-$T$ resistivity manifests itself. This problem has been considered in details in Refs.~[\onlinecite{hwangRMP,Hwang2008, Min2012}] for 2D Dirac materials (i.e. graphene) and in [\onlinecite{Min2015}] for 3D Dirac material.  We will therefore only discuss the high-temperature linear-in-$T$ behavior here for the sake of completeness -- the BG power law temperature dependence of resistivity has been discussed in depth for graphene in Refs.~[\onlinecite{Hwang2008}] and [\onlinecite{Min2012}] and observed experimentally in Ref.~[\onlinecite{EfetovPRL 2010}].

Using the formalism used in Secs. II and III it is easy to show that the high-temperature resistivity in the Dirac systems goes as:
\begin{equation}
\rho (T>T_{BG})=\frac{h}{e^2} \left ( \frac{D}{\hbar v_F v_s} \right )^2 \frac{k_B T}{4 g \rho_m},
\label{rho_gra}
\end{equation}
where $v_F$ is the constant Fermi velocity defining the linear Dirac carrier energy spectrum consisting of $g$ number of equivalent valleys (in graphene $g=2$).  All other quantities ($D$, $v_s$, $\rho_m$) have the same meaning as in sections II and III above.  The Bloch-Gr\"{u}neisen temperature is, as before, $T_{BG}=\hbar v_s k_F/k_B$, where the Fermi wavenumber $k_F$ depends on the carrier density $n$ through $k_F \sim (n/g)^{1/d}$ as before. The crossover temperature $T_c$ or the power law behavior from the high-power to the linear-in-$T$ behavior $T_c$ is again given approximately by $T_c \sim T_{BG}/3$.  Again, we assume that $T_{BG}<T_D$ as appropriate for dilute metals.  It is noteworthy that in contrast to ordinary parabolic systems, the linear-in-$T$ high-temperature resistivity in Dirac systems given by Eq.~(\ref{rho_gra}) is independent of carrier density, as long as $T>T_{BG}$ (with $T_{BG}$ obviously being dependent on the density).

We note that Eq.~(\ref{rho_gra}) implies a very large resistivity if the Fermi velocity $v_F$ of the Dirac material is very small as happens, for example, in the recently discovered twisted bilayer graphene
\cite{cao1,cao2,yank} where $v_F$ could be only a few percent of the graphene Fermi velocity, leading to an extremely high linear-in-$T$ resistivity persisting to low temperatures since the typical carrier density in the twisted bilayer graphene is also small.  This interesting issue has recently been discussed in the literature, pointing out that twisted bilayer graphene is a spectacular example of a strange metal where the strangeness may be arising entirely from ordinary electron-phonon scattering.\cite{wu}

One important difference between parabolic and Dirac systems is that Eq.~(\ref{rho_gra}) describes the temperature dependence of the resistivity in Dirac systems irrespective of whether $T>T_F$ or $T<T_F$ as long as $T>T_{BG}$ (in reality $T>T_c \sim T_{BG}/3$ suffices).  Thus, once the phonon equipartition regime is reached (i.e., $T>T_{BG}$ as we are only considering situations with $T_{BG}<T_D$ appropriate for dilute metals), the linear-in-$T$ resistivity persists in Dirac materials all the way into the nondegenerate regime.  This is apparent already in Eq.~(\ref{rho_gra}) where the right hand side describing the linear T dependence does not contain any quantity which depends on density since $v_F$ in Dirac systems is a density independent constant.

At very high temperatures, with $T$ approaching the characteristic temperature $T_0 \sim T_D$, the phonon phase space restriction comes in because of the Debye cut off in the phonon spectrum, and the  resultant very high-temperature resistivity $\rho(T>T_0)$ starts decreasing with increasing temperature as discussed in sections II and III above for parabolic 2D and 3D systems.  The characteristic  temperature scale $T_0$ for Dirac materials can be shown (following the same procedure as in sec. II and III) to be given by:
\begin{equation}
T_0=\frac{v_F}{g v_s} T_D.   
\end{equation}
In graphene, $T_D \sim 2000$K, and $v_F \gg v_s$, and therefore, this eventual phonon phase space restriction happens at unphysically high temperature ($T>100,000$K) and is of no physical significance.  But if the Debye temperature and/or the Fermi velocity is low in a specific Dirac material, this decrease of $\rho(T)$ below the linear behavior at very high temperatures ($T>T_0$) may become  experimentally observable. Following the methods shown in Sec. II and III, we find the following behavior for $\rho(T)$ at very high temperatures as constrained by the phonon phase space restriction arising from the Debye cut-off:
\begin{eqnarray}
\rho(T) & \sim& T, \;\; {\rm for} T \ll T_0,     \nonumber \\
\rho(T) &\sim & 1/T^2, \;\; {\rm for} T>T_0. 
\end{eqnarray}
Thus, in graphene the suppression of resistivity at very high temperatures, $\rho \sim  T^{-2}$ for $T>T_0$, is much stronger than it is for parabolic 2D materials where [see Eq.~(\ref{eq:rho_td})] $\rho(T) \sim T^{-1/2}$ for $T>T_0$.  We emphasize, however, that for real graphene, with $T_D \sim 2000$K and $v_F/v_s \sim 50$, $T_0 \sim 100,000$K.  Of course, depending on the magnitudes of $T_D$, $v_F$, and $v_s$, there could be Dirac materials where the high-$T$ suppression of $\rho(T)$ imposed by the Debye cut off in the phonon phase space may become relevant.

We note that the linear-in-$T$ resistivity in graphene has already been observed experimentally down to 50K \cite{EfetovPRL 2010} and in very clean graphene the corresponding linearity should persist to arbitrarily low temperatures for lower carrier densities so that $T_{BG}$ is very small.
One possible system where some of the physics being discussed in the current work may become experimentally relevant is twisted bilayer graphene with moire superlattice structure, where the Fermi velocity could be very small near the magic angle,  leading to a possible small value of $T_0$.  In such a system, it is conceivable that the resistivity, after increasing linearly from $T_{BG}$ up to $T_0$, starts decreasing beyond a temperature of order $T_0$, where $T_0$  could be as low as $100-150$K since $v_F$ itself could be very small.\cite{wu}

\section{discussion}

Our results establish that in dilute metals, where $T_{BG}<T_D$ so that the  crossover to the Bloch-Gr\"{u}neisen behavior happens at a rather low temperature $T_c \sim T_{BG}/3$, the phonon scattering induced linear-in-$T$ metallic resistivity could persist to very low temperatures with $T_c \sim n^{1/d}$ being  low for low carrier densities.  In particular, for a 2D system with $n \sim 10^{10}$ cm$^{-2}$, $\rho(T)$ is linear down to $T \sim 1$K whereas at higher densities, e.g., $n \sim 10^{14}$ cm$^{-2}$, $\rho (T)$ is linear down to $\sim$ 50K.  In 3D dilute systems $\rho (T)$, by contrast, manifests a more restricted temperature regime of linear-in-$T$ behavior and the true linearity is apparent only at higher densities where the linearity shows up for $T>50$K.  We emphasize that our definition of a dilute metal is that $T_{BG}<T_D$, i.e., $\hbar k_F v_s < k_B  T_D$, restricting the density range to be ($g$ is the valley degeneracy):
\begin{equation} 
n< (g/2\pi) (k_B T_D/\hbar v_s)^2, \;\; {\rm for \; 2D},  
\end{equation}
\begin{equation}
n< (g/3\pi^2) (k_B T_D/\hbar v_s)^3, \;\; {\rm for \; 3D}.  
\end{equation}
 
In normal metals, $T_D<T_{BG}$, and whereas in most doped 2D structures (including graphene $T_D>T_{BG}$.  Note that typically in a normal metal, $T_F \gg T_{BG} > T_D$, and hence $\rho(T)$ is linear in normal metals roughly from $T_D/3$ up to fairly high temperatures as observed experimentally.  Our results show that in general 2D systems show linear resistivity more strikingly and over a more extended temperature regime than 3D systems.  In particular,  modulation-doped high-mobility (and hence very clean) 2D GaAs structures often experimentally manifest a linear-in-$T$ resistivity down to 1K or below \cite{KawamuraPRB1990} -- in fact, a linear-in-$T$ resistivity down to 50mK (in the temperature range covering two orders of magnitude, $T=0.05-5$K) has been observed in a low-density 2D GaAs system \cite{LillyPRL2003} in agreement with our predictions. In 3D dilute systems by contrast the linear-in-$T$ resistivity is prominent over an extended temperature regime only at higher carrier densities, as shown by our results of Sec.~III.  Of course, any comparison between our theory and experiment depends on the relative magnitude of other resistive scattering mechanisms in the system, but an observation of a linear-in-$T$ resistivity over an extended temperature regime (and down to rather low temperatures) is the rule rather than the exception for a dilute Fermi liquid interacting with acoustic phonons.  There is nothing `strange' about a linear-in-$T$ resistivity persisting to rather low temperatures -- in fact, as our Figs.~3 (for 2D) and 7 (for 3D) show, the linear-in-$T$ resistivity should persist to low temperatures in dilute metals as long as the dominant resistive scattering arises from electron-acoustic phonon interaction.  For 3D systems, in general, the linear-in-$T$ resistivity is more pronounced for larger effective masses.
 
It may be worthwhile to comment on the actual values of $\rho(T)$ due to phonon scattering since our results are presented for the specific parameter sets of GaAs where $D=12$ eV.  For a system with a different deformation potential coupling (but same values of $v_s$ and $T_D$), the resistivity can simply be obtained by multiplying our resistivity results in sections II and III by $(D/12)^2$.  It turns out that most phonon parameters ($D$, $v_s$, $T_D$) do not vary much among electronic materials, and hence our numerical results, although obtained for the GaAs parameters, should apply to most systems of interest as long as the appropriate carrier effective mass and carrier density (which do vary quite a bit among electronic materials with most strongly correlated materials having a low effective carrier density and a large effective carrier mass) are taken into account.  For 2D systems, the resistivity could be as high as $10^6$ ohms ($\sim 40h/e^2$) at room temperatures for a low density of $10^{10}$ cm$^{-2}$ and an effective mass of unity (i.e. free electron mass) whereas at higher densities ($\sim10^{14} $cm$^{-2}$) the resistivity reaches $>10^3$ ohms for unit effective mass.  In 3D systems (again assuming unit effective mass), $\rho (T)$ reaches the colossal magnitude of $10^6$ $\mu \Omega$cm at the low density of $10^{17}$ cm$^{-2}$ and a more reasonable, but still very high, value of 100 $\mu\Omega$cm for a density of $10^{22}$ cm$^{-3}$. Note that the dilute metals have room temperature resistivities much higher than that of normal metals which typically have $\rho \sim 1-2$ $\mu\Omega$cm at room temperatures.  This is simply a reflection of the fact that the resistivity increases essentially linearly with decreasing carrier density, and hence dilute metals could have much larger resistivity than normal metals with their very high carrier densities.  Thus dilute metals have two characteristic qualitative differences with normal metals: (1) The linear-in-$T$ resistivity persists to much lower temperatures by virtue of much lower values of $T_{BG} \sim k_F$, and (2) the actual room temperature  resistivity values are much higher by virtue of $\rho \sim 1/n$ in the zeroth order theory.  Both of these qualitative differences manifest themselves in both 2D and 3D, but the quantitative effect of both is much stronger in 2D.
 We emphasize that our work establishes that both very high electrical resistivity and a linear-in-$T$ resistivity down to rather low temperatures are rather generic features of electron-phonon interaction in dilute metals with low carrier density.  Since the definition of strange or bad metals encompasses precisely these two properties (i.e., linear-in-$T$ resistivity over a large temperature range and rather high resistivity values), the possibility that strange metallic behavior originates from generic electron-phonon interaction operating at carrier densities (effective masses) much lower (higher) than those of normal metals cannot be ruled out.  The fact that most materials regarded as `strange metals' indeed have low carrier densities and high effective masses further reinforces our central claim that strange metallicity may originate from electron-phonon interaction. 
 
We note one important new theoretical finding of our work.  We have discovered a characteristic new temperature scale defined by $T_0 \sim T_D^2/mv_s^2$, which defines a high-temperature scale above which $\rho(T)$ deviates strongly from linearity and in fact, starts decreasing both in 2D and 3D (as well as in Dirac systems). 
This new temperature scale $T_0$ is irrelevant for normal 3D metals where $T_0$ turns out to be unphysically large (100,000K), but in materials with large effective mass and low density, such a $T_0$ may be physically accessible.
In dilute metals, both $T_{BG}<T_D$ and $T_F<T_0$ can be satisfied, leading to a decrease in $\rho(T)$ arising from phonon scattering at very high temperatures, i.e., $\rho(T)$ is increases linearly with $T$  only in the regime $T_{BG}<T<T_0$.  The existence of this characteristic new temperature scale $T_0$, which arises entirely from the phase space restriction in phonon scattering imposed by the Debye momentum cut off, has not been appreciated in the literature before, and it simulates the well-known `resistivity saturation' phenomenon a high temperatures in dilute metals.
Whether such a temperature scale (i.e., $T_0$) is physically accessible in any real material remains an interesting open question for the future.
We note that the experimental `resistivity saturation' is not quite a saturation,\cite{millis}
but more a gentle suppression of $d\rho/dT$ with increasing temperature.  In fact, our calculated results are consistent with such a decreasing $d\rho/dT$ with increasing $T$ at first before an eventual decrease of $\rho(T)$ itself sets in at higher temperatures.  Further work is necessary to establish the physical relevance of our discovered `resistivity saturation' and its connection or not to the experimental resistivity saturation phenomenon.

\section{conclusion}

Our main qualitative conclusion is that the so-called `strange' metallic behavior manifesting in a linear-in-$T$ metallic resistivity over an extended temperature regime (and down to low temperatures) and a very high (compared with normal metals) room-temperature resistivity can both arise generically in a dilute (i.e. low carrier density) 2D or 3D metal due to electron-phonon scattering within the Fermi liquid theory, particularly if the effective mass is also large.  This of course by no means proves or even necessarily suggests the non-existence of non-Fermi-liquid type strange metallic behavior in specific strongly correlated materials since we only show that `diluteness' (low carrier density) is sufficient to produce strange metallicity.  In particular, we cannot rule out other mechanisms (arising, for example, from the proximity to a quantum critical point) giving rise to the strange metallic behavior of extended linear-in-$T$ resistivity and very high resistivity.  Our results do establish, however, that phonon scattering effects should always be considered as a potential source for any strange metallic transport behavior in all low-density metallic systems, but whether phonon scattering is sufficient to explain all of the temperature-dependent resistivity in any particular system depends on the quantitative details which would vary from system to system.  We have, however, emphasized that in low-density high-mobility 2D GaAs structures phonon scattering gives rise to a linear-in-$T$ 2D resistivity which remains linear down to $\sim$ 1K, which at first sight appears very  strange indeed, but can be explained quantitatively by phonon scattering alone without invoking any unknown non-Fermi liquid effects.
 
 We use the semiclassical Boltzmann transport theory within the relaxation time approximation with the scattering rate being calculated in the leading order Fermi's golden rule approximation.  All of these are standard approximations in the literature as long as the electron-phonon coupling is not too large so that higher order scattering contributions remain small.  We restrict ourselves to the dimensionless electron-phonon coupling being less than unity staying within the weak coupling theory so that our approximations remain valid.  Our finding of a resistivity saturation type phenomenon at very high temperatures arises from our treating the phonon phase space restriction correctly taking into account the Debye cut off in theory.  Our high-temperature decrease of resistivity happens at unphysically high temperatures for most normal metals, making it very unlikely that our finding has much to do with the well-known resistivity saturation in resistive metals.  But, for systems with high effective masses and low densities, our predicted high temperature phonon phase space restriction phenomenon may become experimentally observable.  Usually, the standard resistivity saturation phenomenon is studied within a Kubo formula theory.  We are specifically interested in low density metals where the Fermi temperature is low, and hence the neglect of quantum interference effects in our semiclassical Boltzmann theory should not be a bad approximation at high temperatures of interest here.
 
Before concluding, we comment on the currently actively-discussed topic of the so-called `Planckian' dissipative transport\cite{Zhang} in the context of our phonon scattering analysis.  The basic idea of Planckian transport arises from the speculation that many strongly interacting metallic systems are manifestly scale invariant and can be envisioned as a viscous non-Fermi-liquid fluid with a diffusivity bound coming from holographic considerations in string theories.  Since there is no intrinsic scale in such a strongly correlated non-Fermi liquid, the scattering rate must be given by the temperature as the absolute upper bound, and hence the bound is saturated with the assumption that $\hbar/\tau=k_B T$ universally.  This then immediately gives a linear-in-$T$ resistivity from the simple Drude formula:
\begin{equation}
\rho= m/n \tau e^2 \sim T.
\label{eq:48}
\end{equation} 
There is no microscopic physical model for an actual electronic material leading to a theoretical demonstration of such a Planckian dissipation limited resistivity, nevertheless the idea has caught on as a universal description for transport in strange metals, providing a natural, albeit speculative, explanation for $\rho \sim T$ and a particularly large value of $\rho$ since $1/\tau=T$ in this prescription.  In fact, the Planckian dissipation hypothesis has been infused with some empirical support \cite{Bruin} by showing that indeed many electronic materials indeed have a linear-in-$T$ resistivity which provides an effective scattering time $\tau$ with $\hbar/\tau \sim k_B T$ including, rather surprisingly, even many regular normal metals.
 
Our rather modest theory based on electron-phonon interaction in metals, of course, has nothing to do with scale invariance, holography, or Planckian dissipation.  But we do have the (considerable) advantage of being able to calculate the scattering time and the resistivity explicitly starting from a well-known and physically motivated microscopic  model.  In fact, the model we use quantitatively explains the temperature-dependent  resistivity behavior in all normal metals and 2D semiconductor systems (including graphene), obtaining excellent agreement with experimental transport data.  In Fig.~\ref{fig:tau}, we show our calculated $\hbar/\tau$ as a function of $T$ in both 2D and 3D for three different effective mass values ($m^*=$0.1, 1.0, 5.0).  We emphasize that the comparison between $\hbar/\tau$ and $T$ in Fig.~\ref{fig:tau} is an explicitly empirical exercise with no deep significance and is motivated by the empirical analysis of Ref. [\onlinecite{Bruin}] mentioned above.  What we find is rather remarkable in the sense that the finding of Bruin {\it et. al.}\cite{Bruin} is remarkable.  We find that mostly (and always for small effective mass systems) it is indeed true that $\hbar/\tau<k_B T$, but for larger values of effective mass, it is entirely possible for $\hbar/\tau>k_B T$.  Thus, the Planckian dissipative bound can be explicitly violated by electron-phonon scattering depending on the microscopic details.  In fact, once we take into account the fact that the results of Fig.~\ref{fig:tau} are somewhat arbitrary and the calculated scattering rates would change by a factor of $(D/12)^2$ for a different value of the deformation potential coupling (we use $D=12$ eV throughout), we realize that the Planckian dissipation bound is violated by electron-phonon scattering for all strongly coupled electron-phonon systems.  This is of course not surprising at all, but there is perhaps some value in seeing our explicit results presented in Fig.~\ref{fig:tau}.  In particular, we note that purely fortuitously the calculated $\hbar/\tau$ in 3D systems coincides approximately with $k_B T$ for $m^*=1$ around the room temperature as shown in Fig.~\ref{fig:tau}e, providing perhaps a hint that the finding of Ref.~[\onlinecite{Bruin}] may very well be a coincidence.


We expand the above discussion by emphasizing a point not explicitly made in the literature on strange metals and Planckian dissipation.\cite{legros}  It is routinely claimed that a transport scattering rate, $\hbar/\tau$, must necessarily be bounded by $k_B T$ with the bound being saturated by the Planckian limit $\hbar/\tau=k_BT$, which is apparently the maximum possible magnitude for scattering in a metal.  This is, of course, completely incorrect for resistive scattering induced by quenched impurities or acoustic phonons.  For example, the impurity-induced quenched disorder, if sufficiently large in magnitude, could produce a scattering rate much larger than $k_BT$ since impurities simply do not care about the ambient temperature of the carrier system.  In fact, one can always make the system more disordered by adding more impurities, thus arbitrarily increasing $\hbar/\tau$, and thus violating the Planckian limit.   Since quenched disorder produces a finite resistivity (the so-called residual resistivity of metals) even at $T=0$, obviously the claim that $\hbar/\tau < k_BT$ generically fails by definition for impurity scattering.   

More pertinent to our work, it is well-known (and actually in textbooks\cite{grimvall}) that the high-temperature electron-phonon interaction induced scattering rate in metals goes as:
\begin{equation}
\hbar/\tau=2\pi\lambda k_BT,  
\label{eq:49}
\end{equation}
where $\lambda$ is the dimensionless Eliashberg electron-phonon coupling strength.  Putting Eq.~(\ref{eq:49}) in Eq. (\ref{eq:48}) one gets the well-known high-temperature metallic resistivity arising from phonon scattering given by:
\begin{equation}
1/\rho=(\omega_p^2/4\pi)\tau, 
\label{eq:50}
\end{equation}
i.e., 
\begin{equation}
\rho (T)= \frac{8\pi^2\lambda k_BT}{\hbar \omega_p^2}. 
\label{eq:51}
\end{equation}
Equations (\ref{eq:49})--(\ref{eq:51}) apply only in the high-temperature ($T>T_{BG}$ or $T_D$) regime, and $\omega_p$ is the plasma frequency of the metal defined by:
\begin{equation}
\omega_p^2= 4\pi n e^2/m.  
\end{equation}
Our 2D and 3D results presented in the current work are completely consistent with these well-known results, and in fact, using Eq.~(16) we obtain an explicit formula for the 2D dimensionless electron-phonon coupling parameter defined by:
\begin{equation}
\lambda=mD^2/(4\pi \hbar^2 \rho_0 v_s^2),  
\label{eq:53}
\end{equation}
where $\rho_0$ is the 2D mass density.  In general $\lambda_d  \sim (D/v_s \rho_d)^2$, where $\lambda_d$ and $\rho_d$ are respectively the $d$-dimensional dimensionless coupling and atomic mass density.

For the sake of completeness we give below the dimensionless electron-acoustic phonon coupling strength in graphene where the carrier dispersion is linear:
\begin{equation}
\lambda_{\rm graphene}= k_FD^2/(8\pi\hbar v_F \rho_m v_s^2).
\label{eq:54}
\end{equation}
We note that Eqs.~(\ref{eq:53}) and (\ref{eq:54}) are equivalent once we take into account that a density dependent effective mass can be defined for graphene using the identity: $m v_F^2/2=\hbar v_F k_F$, connecting parabolic and linear carrier energy dispersions at the Fermi level (note that the factor 4 difference arises from the chiral property of graphene).  We note (1) the effective coupling for graphene depends on density through $k_F \sim n^1/2$, and (2) if $v_F$ is small, as in twisted bilayer graphene under the flatband condition near the magic angle , the effective electron-phonon coupling would be greatly enhanced.\cite{wu}

Equation (\ref{eq:50}) directly implies that the scattering rate associated with a linear-in-$T$ resistivity could be larger than $k_BT$ if $\lambda>1/2\pi$.  Most metals have $\lambda>1/2\pi$, and thus all metals at room temperatures (or already for $T>50-100$K where the metallic $\rho(T)$ is linear in $T$) strongly violate the so-called Planckian bound!  For example, Al has $\lambda  \sim 0.4$, implying that $\hbar/\tau \sim 2.5 k_BT$, violating the so-called Planckian bound by a factor of 2.5 already for $T>60$K where its resistivity turns linear in $T$.  If most normal metals actually violate the Planckian bound below room temperatures by virtue of Eq.~(\ref{eq:49}), we do not see much value in assigning some special Planckian significance to this bound in the context of defining strange metallicity.  This point is made vividly in our Fig.~\ref{fig:tau} where we show our calculated $\hbar/\tau$ as a function of $T$ for various values of effective mass and carrier density for both 2D and 3D systems. It is clear that for larger values of the effective mass, $\hbar/\tau$ could easily exceed the Planckian bound $k_BT$ at rather low temperatures if the carrier density is not too high!  

We, therefore, do not believe that any significance should be attached to the value of $\tau T$ in transport experiments since both phonon scattering and impurity scattering are capable of producing arbitrarily small values for this dimensionless parameter ($\hbar=k_B=1$ here) depending completely nonuniversally on various system parameters such as impurity density or electron-phonon coupling strength and effective mass and carrier density.  If one knows for certain that the applicable resistive scattering is inelastic arising entirely from the electron-electron interaction, then one can indeed assign special significance to a scattering rate obeying $\hbar/\tau=k_BT$ because it then says something about the underlying Fermi liquid properties of the system since the imaginary part of the self-energy could then be construed to be going as linear-in-$T$ perhaps destroying the underlying Fermi surface, leading to a non-Fermi liquid.  On the other hand, if the scattering time entering resistivity is elastic (as is indeed the case for impurity scattering at all temperatures and for phonon scattering in the linear-in-$T$ equipartition regime of temperatures), its value could be arbitrarily shorter than $1/T$ without implying any non-Fermi liquid behavior.  We do not, however, know how one can be sure whether a measured linear $\rho(T)$ in a particular material in a particular temperature range is arising from inelastic electron-electron scattering using just transport experiments.
We therefore caution against any automatic association of a large scattering rate approaching or surpassing the temperature as being of Planckian in nature -- it is both misleading and incorrect unless one can be absolutely sure that the underlying scattering mechanism is truly inelastic in nature arising indeed from the electron-electron interaction.  Recent experiments on this topic\cite{Bruin,legros} emphasize only the large resistive  scattering rate, without providing any prima facie evidence that this scattering indeed arises from electron-electron interaction, which is necessary to make the case for a non-Fermi liquid strange metal.

 
Finally, we mention very briefly the much-discussed linear-in-$T$ resistivity in some of the cuprates, which originally led to the concept of strange metals.  Cuprates are known to be low-density metals with typical 3D (2D) carrier densities being roughly in the $\sim 10^{21}$ cm$^{-3}$ ($10^{14}$ cm$^{-2}$) depending on the precise value of doping.  
In addition, cuprates also have typically rather large effective masses ($2-10$) and are 2D in nature as far as their transport properties go.  Therefore, cuprates satisfy our minimal criteria for systems where the linear-in-$T$ resistivity could start at rather low temperatures and the resultant resistivity could be rather large (with $\hbar/\tau$ approaching or surpassing $k_{B}T$ because of $2\pi\lambda>1$).
Our calculations indicate that such a low density system could in principle have a linear-in-$T$ resistivity down to $10-50$K depending on the details values of the effective mass, etc.  Depending on the precise value of the effective carrier density and effective mass, the room temperature resistivity could easily be hundreds of $\mu \Omega$cm, but of course, we use a rather simple model without accounting for the details of the cuprates, and studying cuprate transport is well beyond the scope of the current work.
 
To conclude, we show that much of the transport phenomenology of strange metals discussed in the literature can, in principle, arise from the interplay of electron-phonon interaction and low metallic carrier density.  Whether electron-phonon scattering plays an explicit role in the properties of specific strange metals would necessitate detailed quantitative calculations taking into account the microscopic parameters of specific systems, but the possibility that electron-phonon scattering may be playing a role cannot be ruled out {\it a priori}.

{\it Note added}. A recent work by Lavasani, Bulmash, and Das Sarma \cite{ref35} shows that the breakdown of the Wiedemann-Franz law may arise from electron-phonon inter- action, which is consistent with our finding that the linear-in-T resistivity, often associated with the failure of Fermi liquid paradigm, arises also from electron-phonon interaction.

\section*{ACKNOWLEDGMENTS}

This work is supported by Laboratory for Physical Sciences.
E.H.H. also acknowledges support from Basic Science Research Program (2017R1A2A2A05001403) of the National Research Foundation of Korea.


\begin{thebibliography}{99}

\bibitem{Ziman} J. M. Ziman, {\it Principles of the Theory of Solids} (Cambridge, New York, 1972).

\bibitem{Ziman2} J. M. Ziman, {\it Electrons and Phonons} (Oxford, New York, 1960).

\bibitem{ashcroft} N. W. Ashcroft and N. D. Mermin, {\it Solid State Physics} (Saunders, Philadelphia, 1976).

\bibitem{localization0} D. Belitz and T. R. Kirkpatrick, Rev. Mod. Phys. {\bf 66}, 261 (1994).

\bibitem{localization} M. Imada, A. Fujimori, and Y. Tojura (1998), Rev. Mod. Phys. {\bf 70}, 1039 (1998).

\bibitem{localization2} A. Lagendijk, B. van Tiggelen, and D. Wiersma, Phys. Today {\bf 62}, 24 (2009).

\bibitem{ioffe} A. F. Ioffe and A. R. Regel, Prog. Semicond. {\bf 4}, 237 (1960).

\bibitem{mott} N. F. Mott, Philos. Mag. {\bf 26}, 1015 (1972).

\bibitem{kiverson} V. J. Emery and S. A. Kivelson, Phys. Rev. Lett., {\bf 74}, 3253 (1995).

\bibitem{bad} O. Gunnarsson, M. Calandra, and J. E. Han,  Rev. Mod. Phys. {\bf 75}, 1085 (2003).

\bibitem{cuprate} H. Takagi, B. Batlogg, H. L. Kao, J. Kwo, R. J. Cava, J. J. Krajewski, and W. F. Peck Jr., Phys. Rev. Lett. {\bf 69}, 2975 (1992).

\bibitem{ruthenate} S. A. Grigera, R. S. Perry, A. J. Schofield, M. Chiao, S. R. Julian, G. G. Lonzarich, S. I. Ikeda, Y. Maeno, A. J. Millis, and A. P. Mackenzie, Science {\bf 294}, 329 (2001).

\bibitem{titanate} X. Lin, C. W. Rischau, L. Buchauer, A. Jaoui, B. Fauqu\'{e}, and K. Behnia, 
{\it npj} Quantum Materials {\bf 2}, 41 (2017).


\bibitem{manganite} M. B. Salamon and M. Jaime, Rev. Mod. Phys. {\bf 73}, 583 (2001).

\bibitem{heavy} Z. Fisk, D. W. Hess, C. J. Pethick, D. Pines, J. L. Smith, J. D. Thompson, and J. O. Willis, Science, {\bf 239}, 33 (1988).


\bibitem{Bruin} J. A. N. Bruin, H. Sakai, R. S. Perry, and A. P. Mackenzie, Science {\bf 339}, 804 (2013).

\bibitem{zaanen} J. Zaanen, Nature {\bf 430}, 512 (2004).

\bibitem{price} P. J. Price, Ann. Phys. (N.Y.) {\bf 133}, 217 (1981).

\bibitem{KawamuraPRB1990} T. Kawamura and S. Das Sarma, Phys. Rev. B {\bf 45}, 3612 (1992);
Phys. Rev. B {\bf 42}, 3725 (1990).

\bibitem{grimvall} G. Grimvall,  {\it Electron-Phonon Interaction in Metals} (North Holland, Amsterdam, 1981). 

\bibitem{ando} T. Ando, A. B. Fowler, and F. Stern, Rev. Mod. Phys. {\bf 54}, 437 (1982).

\bibitem{hwangRMP} S. Das Sarma, S. Adam, E. H. Hwang, and Enrico Rossi, Rev. Mod. Phys. {\bf 83}, 407 (2011). 

\bibitem{Hwang2008} E. H. Hwang and S. Das Sarma, Phys. Rev. B {\bf 77}, 115449
(2008).


\bibitem{Min2012} H. Min, E. H. Hwang, and S. Das Sarma, Phys. Rev. B {\bf 86} 085307 (2012).

\bibitem{Min2015} S. Das Sarma, E. H. Hwang, and H. Min, Phys. Rev. B {\bf 91}, 035201 (2015).

\bibitem{EfetovPRL 2010} D. K. Efetov and P. Kim, Phys. Rev. Lett. {\bf 105}, 256805 (2010).


\bibitem{cao1} Y. Cao, V. Fatemi, S. Fang, K. Watanabe, T. Taniguchi, E. Kaxiras, and P. Jarillo-Herrero, Nature {\bf 556}, 43 (2018).

\bibitem{cao2} Y. Cao, V. Fatemi, A. Demir, S. Fang, S. L. Tomarken, J. Y. Luo, J. D. Sanchez-Yamagishi, K. Watanabe, T. Taniguchi, E. Kaxiras, R. C. Ashoori, and P. Jarillo- Herrero, Nature {\bf 556}, 80 (2018).

\bibitem{yank}  M. Yankowitz, S. Chen, H. Polshyn, K. Watanabe, T. Taniguchi, D. Graf, A. F. Young, and C. R. Dean, arXiv:1808.07865 (2018).

\bibitem{wu}  F. Wu, E. H. Hwang, S. Das Sarma, arXiv:1811.04920 (2018).


\bibitem{LillyPRL2003} M. P. Lilly, J. L. Reno, J. A. Simmons, I. B. Spielman, J. P. Eisenstein, L. N. Pfeiffer, K. W. West, E. H. Hwang, and S. Das Sarma, Phys. Rev. Lett. {\bf 90}, 056806 (2003).

\bibitem{millis} A. J. Millis, Jun Hu, and S. Das Sarma, Phys. Rev. Lett., {\bf 82}, 2354 (1999).

\bibitem{Zhang} J. Zhang, E. D. Kountz, E. M. Levenson-Falk, R. L. Greene, A. Kapitulnik, arXiv:1808.07564 (2018), and references therein.


\bibitem{legros} A. Legros, S. Benhabib, W. Tabis, F. Laliberté, M. Dion, M. Lizaire, B. Vignolle, D. Vignolles, H. Raffy, Z. Z. Li, P. Auban-Senzier, N. Doiron-Leyraud, P. Fournier, D. Colson, L. Taillefer, and C. Proust, arXiv: 1805.02512 (2018).

\bibitem{ref35} A. Lavasani, D. Bulmash, and S. Das Sarma, Phys. Rev. B {\bf 99}, 085104 (2019).

\end{thebibliography}
\end{document}